\documentclass[times,final]{elsarticle}
\usepackage{jcomp}
\usepackage{framed,multirow}
\usepackage{amssymb,amsmath,amsthm,bm}
\usepackage{latexsym}
\usepackage{mathtools}

\usepackage{subcaption}
\usepackage{graphicx}
\usepackage{diagbox}
\usepackage{booktabs}
\usepackage{makecell}
\usepackage{enumitem} 

\usepackage{url}
\usepackage{xcolor}

\definecolor{newcolor}{rgb}{.8,.349,.1}

\usepackage[mathlines]{lineno}

\newcommand*\patchAmsMathEnvironmentForLineno[1]{
	\expandafter\let\csname old#1\expandafter\endcsname\csname #1\endcsname
	\expandafter\let\csname oldend#1\expandafter\endcsname\csname end#1\endcsname
	\renewenvironment{#1}
	{\linenomath\csname old#1\endcsname}
	{\csname oldend#1\endcsname\endlinenomath}}
\newcommand*\patchBothAmsMathEnvironmentsForLineno[1]{
	\patchAmsMathEnvironmentForLineno{#1}
	\patchAmsMathEnvironmentForLineno{#1*}}
\AtBeginDocument{
	\patchBothAmsMathEnvironmentsForLineno{equation}
	\patchBothAmsMathEnvironmentsForLineno{align}
	\patchBothAmsMathEnvironmentsForLineno{flalign}
	\patchBothAmsMathEnvironmentsForLineno{alignat}
	\patchBothAmsMathEnvironmentsForLineno{gather}
	\patchBothAmsMathEnvironmentsForLineno{multline}
}

\newtheorem{property}{Property}[]

\theoremstyle{definition}

\theoremstyle{remark}
\newtheorem{remark1}{\bf Remark}[]

\begin{document}
	
	\verso{J.~Chen}
	
	\begin{frontmatter}
		
		\title{{\bf Flux-Form Spatiotemporal Neural Operators for Coarse-Grained Dynamics of Multiscale PDEs}}
		
		\author[1]{Junfeng \snm{Chen}}
		\ead{majfchen@ust.hk}
		
		\address[1]{Department of Mathematics, The Hong Kong University of Science and Technology, Hong Kong, China}
		
\begin{keyword}
\KWD \\
Multiscale dynamics\\
Closure modeling\\
Flow map learning\\ 
Neural operators\\
Local conservation
\end{keyword}

\begin{abstract}
We study data-driven prediction of coarse-grained dynamics in multiscale PDE systems. Adopting a closure-free operator-learning viewpoint, we apply a linear coarse-graining map and learn a surrogate evolution operator for the resolved field directly from filtered high-fidelity trajectories. Motivated by the Mori--Zwanzig formalism, we propose a spatiotemporal neural operator mapping a resolved history slab on $\Omega\times[-T_{\mathrm{in}},0]$ to a resolved future slab on $\Omega\times[0,T_{\mathrm{out}}]$. Spatial mixing uses Fourier convolution, while temporal mixing uses a causal kernel operator with position-attention weights on time lags. This causal temporal operator encodes finite-memory effects in the resolved dynamics while preserving the directionality of the history-to-future map. To improve rollout robustness and suppress nonconservative artifacts, we embed a flux-form inductive bias by parameterizing the windowed update in explicit divergence form. We also provide a data-driven guideline for selecting the memory length $T_{\mathrm{in}}$ via the decorrelation time of a closure-injection diagnostic computed from filtered trajectories. We validate on the coarse-grained viscous Burgers' equation, the Kuramoto--Sivashinsky equation, and two-dimensional turbulent flows, obtaining stable autoregressive rollouts with improved long-horizon accuracy and statistical fidelity.

\end{abstract}

\end{frontmatter}


\section{Introduction}\label{sec:intro}
Many systems in science and engineering, including turbulent flows, climate dynamics, reactive transport, and multiphase phenomena, all exhibit pronounced multiscale behavior. Even when governing equations are known, resolving all relevant scales by direct numerical simulation is often computationally prohibitive. This motivates \emph{coarse-grained} models that evolve only resolved variables, obtained for instance by filtering or projection. A central obstacle is the classical \emph{closure problem}~\cite{duraisamy2019turbulence,sanderse2025scientific}: after coarse-graining, the resolved dynamics are generally not closed because they depend on unresolved degrees of freedom. In fluid mechanics this appears as subgrid-scale modeling in large-eddy simulation and stress closures in Reynolds-averaged models. While traditional closures built on first principles and empirical knowledge have enabled major progress, their structural assumptions and regime-specific calibration can limit their reliability beyond the calibrated regime.

Machine learning has recently accelerated data-driven closure modeling. A dominant line of work augments a prescribed resolved PDE with an \emph{in-equation} correction term and learns that term from data, either by supervised regression on closure targets (a priori)~\cite{parish2016paradigm,ling2016reynolds,wang2017physics,huang2022machine} or by end-to-end, solver-in-the-loop training that optimizes rollout behavior (a posteriori)~\cite{sirignano2020dpm,macart2021embedded,list2022learned,melchers2023comparison,boral2023neural,van2024energy}. Modern approaches incorporate structures like symmetries~\cite{huang2023machine1,huang2023machine2}, physical bounds~\cite{freeman2024quantum}, and nonlocality~\cite{pan2018data,ma2019model,beck2019deep,wang2020recurrent,guan2022stable,guan2023learning}. Despite these advances, in-equation strategies typically commit to a particular and rigid PDE form, and their long-horizon behavior can be sensitive to solver coupling, numerical errors, and distribution shift across regimes.

An alternative paradigm is \emph{equation-free} coarse-grained modeling: rather than identifying an explicit closure term, one learns the evolution of the resolved variables directly from data. This includes (i) reduced-coordinate modeling in latent spaces obtained by POD or autoencoders~\cite{maulik2020time}, and (ii) field-level operator learning with neural operators that approximate mappings between function spaces~\cite{li2021fourier,li2022fourier,wang2024beyond}. Equation-free approaches avoid prescribing a closure formula, but they do not by themselves solve the closure problem. After coarse-graining, the resolved dynamics are generally history-dependent, and the unresolved feedback must still respect the conservative structure of the governing equations. Without these inductive biases, learned flow maps may achieve accurate short-term interpolation while producing unstable or statistically inconsistent long-horizon rollouts.

The Mori--Zwanzig (MZ) formalism makes the source of these difficulties explicit. Projecting a Markovian full system onto resolved variables yields a generalized Langevin equation consisting of a Markov term, a memory integral over the entire past, and an orthogonal-dynamics fluctuation term~\cite{mori1965transport,zwanzig1973nonlinear}. This shows that coarse-grained dynamics are not closed by an instantaneous state alone: unresolved scales feed back through history-dependent and fluctuation  effects. When the memory contribution is effectively localized in time, it is natural to approximate this feedback by a finite recent history~\cite{chorin2000optimal,chorin2002optimal,stinis2004stochastic,lu2017data,wang2020recurrent,fu2020learning}. We therefore model coarse-grained prediction as a causal finite-memory history-to-future map, rather than as a Markovian one-step update. The resulting operator is not intended to identify the exact MZ memory kernel; it is a data-driven surrogate for the effective resolved evolution induced by unresolved scales.

In this work, we develop an equation-free approach for coarse-grained multiscale PDE dynamics that is aligned with the truncated-memory MZ viewpoint and tailored to stable long-horizon rollout. We treat the recent resolved trajectory as an extended state: a history slab on $\Omega\times[-T_{\mathrm{in}},0]$ is mapped to a future slab on $\Omega\times[0,T_{\mathrm{out}}]$ by a learned windowed evolution operator. The model is a spatiotemporal neural operator with structured space--time mixing. Spatial interactions are represented by Fourier convolution~\cite{li2021fourier}, providing a translation-equivariant spectral parameterization of resolved fields. Temporal interactions are represented by a causal kernel operator built from position-attention weights~\cite{chen2024positional} and an explicit causal mask. This construction encodes the finite-memory and time-directed character of coarse-grained dynamics at the operator level.

Beyond operator parameterization, we impose a conservative inductive bias through the flux-form structure of the underlying PDE. Instead of regressing the future resolved field as an unconstrained output, the network predicts an effective flux over the output window, and the resolved state is updated by an explicit divergence-form rule anchored at the last observed snapshot. This parameterization enforces local conservation by construction: changes of resolved mass occur only through boundary fluxes; this suppresses nonconservative drift during autoregressive rollout. Finally, instead of treating the memory horizon as a free architecture parameter, we select $T_{\mathrm{in}}$ using a data-driven decorrelation time estimated from an effective closure-injection diagnostic computed from filtered trajectories.

Our main contributions are:
\begin{itemize}
\item \textbf{Equation-free finite-memory forecasting for coarse-grained multiscale dynamics.} 
We formulate coarse-grained prediction as learning a windowed operator from a resolved history slab to a resolved future slab, consistent with truncated-memory interpretations of coarse-graining-induced non-Markovianity.
\item \textbf{Causal temporal operator for history-dependent reduced dynamics.} 
We introduce a causal temporal kernel based on position-attention with an explicit causal mask, providing a time-directed mechanism for mixing resolved histories within a spatiotemporal neural-operator architecture.
\item \textbf{Flux-form inductive bias for locally conservative rollout.} 
We parameterize the learned evolution through an explicit divergence-form update anchored at the last resolved snapshot, enforcing local conservation by construction and reducing nonconservative drift under autoregressive rollout.
\item \textbf{Data-driven memory-length selection.} 
We provide a practical guideline for selecting $T_{\mathrm{in}}$ via the decorrelation time of an effective closure-injection diagnostic computed from filtered trajectories.
\item \textbf{Canonical multiscale PDE benchmarks and ablations.} 
We validate the method on spectrally coarse-grained viscous Burgers', Kuramoto--Sivashinsky, and two-dimensional Navier--Stokes dynamics, compare against representative equation-free and physics-based baselines, and ablate the roles of the flux-form parameterization and the memory length $T_{\mathrm{in}}$.
\end{itemize}

The remainder of the paper is organized as follows. Section~\ref{sec:setup} introduces coarse-graining and the closure problem, reviews the MZ generalized Langevin equation, and motivates truncated-memory modeling. Section~\ref{sec:methods} presents the spatiotemporal neural operator, causal temporal kernel construction, flux-form update rule, and memory-length estimation procedure. Section~\ref{sec:experiments} reports numerical results and ablations on Burgers', Kuramoto--Sivashinsky, and Navier--Stokes benchmarks. Section~\ref{sec:discussions} concludes with limitations and outlook.

\section{Setup and preliminaries}
\label{sec:setup}

\subsection{Coarse-graining and the closure problem}\label{sec:setup_closure}
We consider PDEs in the following conservative form:
\begin{equation}\label{eq:general_pde}
u_t + \nabla\cdot\,F\left(u, \nabla u, \nabla^2u,\dots\right) = 0,
\end{equation}
where $u$ is an $\mathbb{R}^{d_u}-$valued function defined on $\Omega\times [0,T]\subset\mathbb{R}^d\times\mathbb{R}^+$, $\nabla\cdot$ is the divergence operator, $F(\cdot)$ denotes a \emph{known} generic \emph{nonlinear flux} function which can depend on solution $u$ and its spatial derivatives. This formulation covers many PDEs of practical interest. For example, the flux for one-dimensional viscous Burgers' equation is 
$$\frac{1}{2}u^2-\nu\,u_{x},$$
which is composed of two terms standing respectively for convection and diffusion.
For one-dimensional Kuramoto--Sivashinsky equation, the flux can be written as
$$\frac{1}{2}u^2+u_{x}+u_{xxx},$$
where the third term is related with dispersion. Similar organization applies to two-dimensional incompressible Navier--Stokes equations with conservative external forcing, which can be written in the vorticity form as
$$\omega_t + \nabla\cdot\,\left(\omega \mathbf{v}-\nu\nabla\omega-\mathbf{f}\right)=0,$$
where $\mathbf{v}$ is the velocity, $\omega=\nabla\times\mathbf{v}$ denotes the vorticity, and $\mathbf{f}$ is the external forcing.

Let $\mathcal{P}$ denote a \emph{linear} coarse-graining operator (e.g., a sharp spectral projector, spatial filter, or other bounded linear projection) and define the resolved field
\begin{equation}\label{eq:resolved_def}
\bar u = \mathcal{P}u.
\end{equation}
Applying $\mathcal{P}$ to \eqref{eq:general_pde} yields the projected dynamics
\begin{equation}\label{eq:proj_dynamics}
\bar u_t + \nabla\cdot\,\mathcal{P}F(u, \nabla u, \nabla^2u,\dots) = 0.
\end{equation}
The difficulty is that $\nabla\cdot\,\mathcal{P}F(u, \nabla u, \nabla^2u,\dots)$ depends on the full state $u$, not solely on $\bar u$, hence \eqref{eq:proj_dynamics} is not closed in the resolved variables.

A common approximation replaces the unclosed term $\nabla\cdot\,\mathcal{P}F(u, \nabla u, \nabla^2u,\dots)$ by $\nabla\cdot\,F(\bar u, \nabla\bar u, \nabla^2\bar u,\dots)$, which leads to the residual (closure forcing)
\begin{equation}\label{eq:closure_residual}
r(x,t) := \nabla\cdot\,\Big(F(\bar u, \nabla\bar u, \nabla^2\bar u,\dots)-\mathcal{P}F(u, \nabla u, \nabla^2u, \dots)\Big).
\end{equation}
Equivalently, the resolved evolution can be written as
\begin{equation}\label{eq:resolved_with_residual}
\bar u_t + \nabla\cdot\,F(\bar u, \nabla\bar u, \nabla^2\bar u,\dots) = r.
\end{equation}
The field $r$ aggregates the net effect of unresolved scales on the resolved dynamics. Therefore, without querying the full state $u$, the true evolution law for $\bar u$ is \emph{unknown}, and our ultimate goal is to construct a predictive model for $\bar u$'s dynamics from available snapshot data
\begin{equation}\label{eq:data_general}
    {\bar u}^{(i)}_0, {\bar u}^{(i)}_1,\dots,{\bar u}^{(i)}_K,\qquad i=1,2,\dots,I,
\end{equation}

where $I\ge 1$ is the number of available trajectories, and ${\bar u}^{(i)}_k$ denotes the resolved state at time $t=k\Delta$ on the $i$th trajectory. The snapshot data are obtained after applying the coarse-graining operator $\mathcal{P}$ to high-fidelity solutions of \eqref{eq:general_pde}, and are represented on the spatial and temporal grids used for training. For a fixed full equation \eqref{eq:general_pde} and a fixed coarse-graining operator $\mathcal{P}$, the object of interest is the induced resolved evolution of $\bar u$. Our goal is to learn a closure-free surrogate for this induced resolved evolution from filtered trajectory data, while later imposing finite-memory and conservative structure to improve long-horizon rollout.

\subsection{Mori--Zwanzig formalism}\label{sec:setup_mz}
The Mori--Zwanzig (MZ) formalism~\cite{mori1965transport,zwanzig1973nonlinear} provides an exact representation of coarse-grained dynamics and makes explicit that projection generally induces memory. In particular, for a resolved variable $\bar u(t)$ associated with a projection $\mathcal{P}$, the resolved evolution can be written in the generalized Langevin equation (GLE) form
\begin{equation}\label{eq:gle}
\frac{d}{dt}\bar u(t)
=
\underbrace{\mathcal{R}\!\left(\bar u(t)\right)}_{\text{Markov term}}
+
\underbrace{\int_{0}^{t}\mathcal{K}\!\left(s;\bar u(t-s)\right)\,ds}_{\text{memory term}}
+
\underbrace{\eta(t)}_{\text{noise / orthogonal dynamics}}.
\end{equation}
Equation \eqref{eq:gle} decomposes the exact resolved dynamics into three contributions: 
(i) a Markov term $\mathcal{R}(\bar u(t))$ depending on the instantaneous resolved state; 
(ii) a memory term given by a convolution against a kernel $\mathcal{K}$ that couples the present rate of change to the entire past history $\{\bar u(t-s)\}_{0\le s\le t}$; and 
(iii) a fluctuation term $\eta(t)$ generated by the orthogonal dynamics and depending on the unresolved components. The presence of the history integral implies that the resolved dynamics are generically non-Markovian, which should apply to the closure forcing field $r(x,t)$ (see equations \eqref{eq:closure_residual}, \eqref{eq:resolved_with_residual}).

\subsection{Truncated-memory approximations}\label{sec:setup_truncated_memory}
In practical reduced modeling, the full history dependence in \eqref{eq:gle} is typically approximated by a truncated-memory model. Two closely related viewpoints appear in the literature.

\paragraph{Short-memory truncation}
If the memory kernel $\mathcal{K}(s;\cdot)$ decays sufficiently fast with the lag $s$, one may approximate the history integral in \eqref{eq:gle} by retaining only a recent window of memory length $T_{\mathrm{in}}$,
\begin{equation}\label{eq:finite_memory}
\int_{0}^{t}\mathcal{K}\!\left(s;\bar u(t-s)\right)\,ds
\;\approx\;
\int_{0}^{T_{\mathrm{in}}}\mathcal{K}\!\left(s;\bar u(t-s)\right)\,ds,
\end{equation}
thereby truncating the dependence on the distant past. For many real-world physical systems, this approximation is effective~\cite{chorin2000optimal,chorin2002optimal,stinis2004stochastic,lu2017data,wang2020recurrent,fu2020learning}, as the past unresolved dynamics eventually decorrelate from the resolved evolution. One may need longer memory length for systems without a strong scale decorrelation~\cite{stinis2015renormalized,price2019renormalized}.

\paragraph{On the noise term}
The fluctuation term $\eta(t)$ in \eqref{eq:gle} is unclosed in general, so the map from resolved history to resolved future is generally not deterministic at the level of $\bar u$ alone: different realizations of the unresolved degrees of freedom can lead to different futures even when the resolved history is similar. However, deterministic reduced models frequently neglect the noise or absorb it into an effective memory approximation, which can be understood as the mean of unresolved fluctuations conditioned on the resolved history.

\section{Equation-free learning framework}
\label{sec:methods}
In this work we adopt the standard deterministic, truncated-memory viewpoint: we neglect the noise term and approximate the memory integral by a finite horizon $T_{\mathrm{in}}$, leading formally to
\begin{equation}\label{eq:gle_truncated}
\frac{d}{dt}\bar u(t)
\;=\;
\mathcal{R}\!\left(\bar u(t)\right)
+
\int_{0}^{T_{\mathrm{in}}}\mathcal{K}\!\left(s;\bar u(t-s)\right)\,ds,
\end{equation}
Direct numerical treatment of~\eqref{eq:gle_truncated} requires evaluating a history integral (and, in general, solving a delay-type system) at every time step, which is nontrivial in high-dimensional spatiotemporal settings.

From a modeling perspective, we do \emph{not} attempt to identify the Markovian term $\mathcal{R}$ or the memory kernel $\mathcal{K}$. Instead, we adopt an equation-free strategy: we learn a coarse-grained evolution operator directly from snapshot data generated by the full system~\eqref{eq:general_pde}. The objective is a surrogate flow map~\cite{qin2019data,fu2020learning,churchill2023flow} that advances the resolved field while accounting for coarse-graining-induced non-Markovianity.

\subsection{Space--time operator learning with memory}\label{sec:methods_stno}
Fix horizons $T_{\mathrm{in}},T_{\mathrm{out}}>0$. For each reference time $t$, define the history and future slabs
\begin{equation}\label{eq:slabs_def_methods}
    \begin{aligned}
        \bar u^0_{-T_{\mathrm{in}}}(x,\tau) &:= \bar u(x,t+\tau),\quad \tau\in[-T_{\mathrm{in}},0],\\
        \bar u_0^{T_{\mathrm{out}}}(x,\tau) &:= \bar u(x,t+\tau),\quad \tau\in[0,T_{\mathrm{out}}],
    \end{aligned}
\end{equation}
viewed as elements of function spaces 
\[
\mathcal{X} := L^1\!\big([-T_{\mathrm{in}},0];\,L^2(\Omega)\big),\qquad
\mathcal{Y} := L^1\!\big([0,T_{\mathrm{out}}];\,L^2(\Omega)\big),
\]
with norms
\[
\|v\|_{\mathcal{X}}
:= \frac{1}{T_{\mathrm{in}}}\int_{-T_{\mathrm{in}}}^{0}\|v(\cdot,\tau)\|_{L^2(\Omega)}\,d\tau,
\qquad
\|w\|_{\mathcal{Y}}
:= \frac{1}{T_{\mathrm{out}}}\int_{0}^{T_{\mathrm{out}}}\|w(\cdot,\tau)\|_{L^2(\Omega)}\,d\tau.
\]
We then seek an operator $\mathcal{G}:\mathcal{X}\to\mathcal{Y}$ mapping a resolved history segment to a resolved future segment.

Following standard neural-operator constructions~\cite{kovachki2023neural,chen2025due}, we begin with trainable \emph{linear} integral operators acting jointly over space and time. Given a function $v:\Omega\times[t_1,t_2]\to\mathbb{R}^{d_v}$, we define
\begin{equation}\label{eq:st_integral}
(\mathcal{T}v)(x,t)
=
\int_{t_1}^{t_2}\int_{\Omega}
\kappa(x,t;y,s)\,v(y,s)\,\mathrm{d}y\,\mathrm{d}s,
\end{equation}
where $\kappa$ is a learnable kernel and $[t_1,t_2]$ is a prescribed time interval. The operator $\mathcal{T}$ mixes information across spatial locations and across time.

A spatiotemporal neural operator is constructed by composing such integral layers with pointwise nonlinearities. In abstract form, for $\ell=0,\dots,L-1$,
\begin{equation}\label{eq:stno_abstract}
v^{(0)} = \mathcal{E}\big(\bar u^0_{-T_{\mathrm{in}}}\big),\qquad
v^{(\ell+1)} = \sigma\!\left(\mathcal{W}^{(\ell)}v^{(\ell)} + \mathcal{T}^{(\ell)}v^{(\ell)}\right),\qquad
\widehat{\bar u}_0^{\mathrm{out}} = \mathcal{D}\big(v^{(L)}\big),
\end{equation}
where $v^{(\ell)}: \Omega\times[t_1^{(\ell)},t_2^{(\ell)}]\to\mathbb{R}^{d^{(\ell)}}$ denotes an intermediate space--time function defined on an intermediate time interval $[t_1^{(\ell)},t_2^{(\ell)}]$, $\mathcal{E}: \mathbb{R}_{d_{\bar u}}\to\mathbb{R}^{d^{(0)}}$ and $\mathcal{D}: \mathbb{R}^{d^{(L)}}\to\mathbb{R}^{d_{\bar u}}$ are finite-dimensional local maps (typically trainable linear layers), $\mathcal{W}^{(\ell)}: \mathbb{R}^{d^{(\ell)}}\to\mathbb{R}^{d^{(\ell+1)}}$ are also finite-dimensional local maps for channel mixing purpose, and $\sigma$ is an elementwise nonlinearity. This construction treats the recent resolved trajectory as an extended state and learns a windowed history-to-future evolution operator instead of an instantaneous Markovian update.

On dense space--time grids, learning an unrestricted kernel $\kappa(x,t;y,s)$ is computationally prohibitive. We therefore adopt a structured parameterization in which the space--time kernel factorizes into a spatial and a temporal component,
\begin{equation}\label{eq:space_time_kernel}
\kappa(x,t;y,s)=\kappa^{\mathrm{space}}(x-y)\,\kappa^{\mathrm{time}}(t,s).
\end{equation}
Accordingly, each $\mathcal{T}$ is implemented as a spatial convolution followed by a temporal operator that maps from one time grid to another. For the spatial component, we use standard Fourier convolution, which implements spatial convolution as multiplication in frequency space and is translation-equivariant. For the temporal component, although it is numerically possible to apply Fourier convolution by treating time as an additional coordinate~\cite{cao2025spectral}, Fourier mixing over the time axis is inherently two-sided and does not encode causal directionality. The next subsection introduces a causal temporal operator tailored to the history-to-future setting.

\subsection{Temporal convolution respecting causality}\label{sec:methods_causal_conv}

For spatiotemporal operator learning, a crucial requirement is that the learned operator be \emph{causal in time}: predictions at a query time may depend only on inputs from earlier (or equal) times. Causality is not automatic for a generic operator of the form~\eqref{eq:st_integral}. We enforce causality by constructing the temporal component of $\mathcal{T}$ as a masked temporal convolution on time nodes.

Let $I_{\mathrm{in}}=[t_1,t_2]$ and $I_{\mathrm{out}}=[\tilde t_1,\tilde t_2]$ denote two time intervals. Temporal mixing is implemented as
\begin{equation}\label{eq:temporal_integral}
(\mathcal{T}^{\mathrm{time}}v)(x,\tilde t)
=
\int_{I_{\mathrm{in}}} \kappa^{\mathrm{time}}(\tilde t,t)\,v(x,t)\,dt,\qquad \tilde t \in I_{\mathrm{out}},
\end{equation}
with a trainable kernel $\kappa^{\mathrm{time}}(\tilde t,t)$. Causality corresponds to the support constraint
\begin{equation}\label{eq:causal_support}
\kappa^{\mathrm{time}}(\tilde t, t)=0\qquad \text{whenever}\quad t>\tilde t,
\end{equation}
so that $(\mathcal{T}^{\mathrm{time}}v)(x,\tilde t)$ depends only on admissible past times.

We parameterize $\kappa^{\mathrm{time}}$ using attention weights defined purely by temporal positions, following the position-attention mechanism in~\cite{chen2024positional}. Adapting squared-distance attention to time and incorporating causal masking yields
\begin{equation}\label{eq:time_posatt}
\kappa^{\mathrm{time}}(\tilde t,t)
=
\frac{\exp\!\big(-\lambda(\tilde t-t)^2\big)\,\mathbf{1}_{t\le \tilde t}}
{\int_{I_{\mathrm{in}}}\exp\!\big(-\lambda(\tilde t-t')^2\big)\,\mathbf{1}_{t'\le \tilde t}\,\mathrm{d}t'},
\end{equation}
where $\lambda\ge 0$ is a trainable scale parameter. Relative to position-attention in space, the novelty here is the causal masking, which converts a symmetric distance-based kernel into a valid temporal kernel for history-dependent prediction. Such an embedded time causality distinguishes our approach from spatiotemporal Fourier mixing formulations that treat time as a coordinate without enforcing causal support~\cite{cao2025spectral}. Causality here is a structural constraint rather than a performance device, as it makes the history-to-future operator well-posed and interpretable as a non-Markovian resolved dynamics model.

Let $\{t_i\}_{i=1}^{m}\subset I_{\mathrm{in}}$ denote equispaced input time nodes. For any output query $\tilde t\in I_{\mathrm{out}}$, we approximate~\eqref{eq:temporal_integral} by the normalized quadrature rule
\begin{equation}\label{eq:time_conv_discrete}
(\mathcal{T}^{\mathrm{time}}v)(x,\tilde t)
\;\approx\;
\sum_{i=1}^{m} w_i(\tilde t)\,v(x,t_i),
\qquad
w_i(\tilde t):=
\frac{\exp\!\big(-\lambda(\tilde t-t_i)^2\big)\,\mathbf{1}_{t_i\le \tilde t}}
{\sum\limits_{i'=1}^{m}\exp\!\big(-\lambda(\tilde t-t_{i'})^2\big)\,\mathbf{1}_{t_{i'}\le \tilde t}}.
\end{equation}

As in Equation \eqref{eq:stno_abstract}, we allow intermediate time grids to change across layers. In practice we use
\begin{equation}\label{eq:time_intervals}
     \begin{aligned}
     [t_1^{(0)},t_2^{(0)}] &= [-T_{\mathrm{in}}, 0], \\
     [t_1^{(\ell)},t_2^{(\ell)}] &= [-T_{\mathrm{in}}, T_{\mathrm{out}}],\quad \ell=1,\dots,L-1, \\
     (t_1^{(L)},t_2^{(L)}] &= [0, T_{\mathrm{out}}].
     \end{aligned}
\end{equation}
so the first temporal layer maps the input history slab to an extended latent interval, intermediate blocks operate on this latent interval, and the final temporal layer maps to the target interval.

\begin{remark1}[Connection to kernel regression and expressivity]
For fixed $\tilde t$, the weights $w_i(\tilde t)$ in~\eqref{eq:time_conv_discrete} define a causal Nadaraya--Watson estimator~\cite{nadaraya1964estimating,watson1964smooth}. In contrast to classical kernel regression where the bandwidth is selected a priori, here $\lambda$ is learned, yielding a data-adaptive temporal receptive field. Moreover, \eqref{eq:time_conv_discrete} represents only a single linear smoothing operator; stacking such temporal operators with spatial convolution and nonlinear activations (Section~\ref{sec:methods_stno}) yields substantially higher expressive capacity than a single kernel smoother.
\end{remark1}

\begin{remark1}
    It is useful to distinguish our windowed history-to-future operator learning from approaches that approximate a full solution map over a long time interval by treating time as an input coordinate (e.g., physics-informed neural networks~\cite{raissi2019physics}). Here $\bar u^0_{-T_{\mathrm{in}}}$ is an extended state, and $\mathcal{G}$ advances this state over a fixed horizon $T_{\mathrm{out}}$. Long-horizon forecasting is obtained by composition of this short-horizon map, rather than by directly querying the model at arbitrarily large times. Consequently, the relevant notion of generalization is robustness with respect to the distribution of history slabs encountered during autoregressive rollout.
\end{remark1}
\subsection{Embedding a flux-form inductive bias}\label{sec:methods_increment}
As discussed in Section~\ref{sec:setup_closure}, many PDEs of interest admit a conservative (divergence) form
\[
u_t+\nabla\cdot F(u,\nabla u,\nabla^2u,\dots)=0.
\]
For coarse-grained dynamics, the projected equation~\eqref{eq:resolved_with_residual}
introduces an unknown forcing $r$ that aggregates unresolved effects. We embed a flux-form inductive bias by
representing this forcing in divergence form. Specifically, rewriting~\eqref{eq:closure_residual} yields
\begin{equation}\label{eq:r_as_div_phi}
r(\cdot,t)
=\nabla\cdot\Big(F_{\mathrm{phys}}(\bar u(\cdot,t))-\mathcal{P}F(u(\cdot,t),\nabla u(\cdot,t),\dots)\Big)
=: \nabla\cdot\,\phi(\cdot,t),
\end{equation}
where $F_{\mathrm{phys}}(\bar u):=F(\bar u,\nabla\bar u,\nabla^2\bar u,\dots)$ denotes the resolved flux and
$\phi$ is an unknown flux-like field capturing closure effects.

Equation~\eqref{eq:r_as_div_phi} motivates an explicit flux-form ansatz for the windowed evolution operator G: for $\tilde t\in[0,T_{\mathrm{out}}]$,
\begin{equation}\label{eq:update_rule}
\bar u(\cdot,\tilde t) = \mathcal{G}(\bar u^0_{-T_{\mathrm{in}}})(\cdot,\tilde t)
=
\bar u(\cdot,0)
-\tilde t\,\nabla\cdot\Big(\Psi\!\big(\bar u^0_{-T_{\mathrm{in}}}\big)(\cdot,\tilde t)\Big),
\end{equation}
Equation~\eqref{eq:update_rule} is a \emph{window-level parameterization} of the resolved evolution operator, rather than an identification of a pointwise-in-time closure term. For each fixed history slab $\bar u^0_{-T_{\mathrm{in}}}$ and each $\tilde t\in[0,T_{\mathrm{out}}]$, one may view $\Psi(\bar u^0_{-T_{\mathrm{in}}})(\cdot,\tilde t)$ as an \emph{effective flux} that makes the conservative-form relation~\eqref{eq:update_rule} hold for the observed increment $\bar u(\cdot,0)\mapsto \bar u(\cdot,\tilde t)$. The explicit factor $\tilde t$ enforces the anchoring condition $\bar u(\cdot,0)=\mathcal{G}(\bar u^0_{-T_{\mathrm{in}}})(\cdot,0)$. Moreover, $\Psi$ is not uniquely defined: only its divergence-relevant component affects~\eqref{eq:update_rule}, so any divergence-free component is immaterial. Related conservation-encoding perspectives in learned dynamics and neural operators have also been explored in~\cite{richter2022neural,liu2024harnessing}.

\paragraph{Optional physics-informed decomposition (frozen resolved flux)}
When a reliable expression for the resolved flux $F_{\mathrm{phys}}$ is available and inexpensive to evaluate, one may
optionally decompose the effective flux as
\[
\Psi\!\big(\bar u^0_{-T_{\mathrm{in}}}\big)(\cdot,\tilde t)
=
F_{\mathrm{phys}}(\bar u(\cdot,0))-\Psi'\!\big(\bar u^0_{-T_{\mathrm{in}}}\big)(\cdot,\tilde t),
\qquad \tilde t\in[0,T_{\mathrm{out}}],
\]
This choice can be advantageous when $F_{\mathrm{phys}}(\bar u(\cdot,0))$
captures the dominant resolved transport over the window (e.g., moderate $T_{\mathrm{out}}$ or slowly varying resolved
states), since it reduces the learning burden by focusing the network on closure effects and often improves
data-efficiency and out-of-distribution robustness.

The learning task is therefore to approximate such an effective representation by $\widehat{\Psi}$, implemented as a spatiotemporal neural operator. The resulting predictor is obtained by substituting
$\widehat{\Psi}$ into Equation~\eqref{eq:update_rule}:
\begin{equation}\label{eq:update_rule_hat}
\widehat{\bar u}(\cdot,\tilde t)
=
\bar u(\cdot,0)
-\tilde t\,\nabla\cdot\Big(\widehat{\Psi}\!\big(\bar u^0_{-T_{\mathrm{in}}}\big)(\cdot,\tilde t)\Big),
\qquad \tilde t\in[0,T_{\mathrm{out}}].
\end{equation}

We next record two immediate consequences of this conservative-form parameterization, which motivate
the design~\eqref{eq:update_rule_hat}.

\begin{property}[Local conservation]\label{prop:local_conservation}
For any measurable subdomain $D\subset\Omega$ with sufficiently regular boundary $\partial D$ and outward unit normal
$n$, the predictor~\eqref{eq:update_rule_hat} satisfies, for all $\tilde t\in[0,T_{\mathrm{out}}]$,
\begin{equation}\label{eq:local_conservation}
\int_D \widehat{\bar u}(x,\tilde t)\,dx
=
\int_D \bar u(x,0)\,dx
-\tilde t\int_{\partial D}\widehat{\Psi}(\bar u^0_{-T_{\mathrm{in}}})(x,\tilde t)\cdot n\,dS.
\end{equation}
In particular, if $D=\Omega$ and the boundary conditions are periodic or no-flux, then
$\int_\Omega \widehat{\bar u}(x,\tilde t)\,dx=\int_\Omega \bar u(x,0)\,dx$.
\end{property}

\begin{proof}
Integrate~\eqref{eq:update_rule_hat} over $D$ and apply the divergence theorem to obtain
\[
\int_D \nabla\cdot\Big(\widehat{\Psi}(\bar u^0_{-T_{\mathrm{in}}})(\cdot,\tilde t)\Big)\,dx
=
\int_{\partial D}\widehat{\Psi}(\bar u^0_{-T_{\mathrm{in}}})(x,\tilde t)\cdot n\,dS.
\]
Substituting into~\eqref{eq:update_rule_hat} gives~\eqref{eq:local_conservation}. The final claim follows by taking
$D=\Omega$ and using the boundary conditions to cancel the boundary flux integral.
\end{proof}

Property~\ref{prop:local_conservation} shows that conservative structure is enforced by construction: the predicted
mass in any subdomain changes only through boundary fluxes. This is the main qualitative difference between the
flux-form predictor~\eqref{eq:update_rule_hat} and generic black-box window-to-window regressors.

To formalize how the conservative form shapes the training signal across Fourier modes, it is convenient to note that
for each fixed trajectory,~\eqref{eq:update_rule} identifies only $\nabla\cdot\Psi$ (any divergence-free component is
immaterial). Rearranging~\eqref{eq:update_rule} gives, for $\tilde t\in[0,T_{\mathrm{out}}]$,
\begin{equation}\label{eq:phi_defined_by_increment}
\nabla\cdot\Big(\Psi(\bar u^0_{-T_{\mathrm{in}}})(\cdot,\tilde t)\Big)
=\frac{\bar u(\cdot,0)-\bar u(\cdot,\tilde t)}{\tilde t}.
\end{equation}

\begin{property}[Mode-weighting induced by conservative-form parameterization]\label{prop:mode_weighting}
Let $\Psi(\bar u^0_{-T_{\mathrm{in}}})(\cdot,\tilde t)$ satisfy~\eqref{eq:update_rule} (equivalently~\eqref{eq:phi_defined_by_increment})
for the resolved data, and define the mismatches
\[
\delta\psi(\cdot,\tilde t):=\widehat{\Psi}(\bar u^0_{-T_{\mathrm{in}}})(\cdot,\tilde t)-\Psi(\bar u^0_{-T_{\mathrm{in}}})(\cdot,\tilde t),
\qquad
\delta u(\cdot,\tilde t):=\widehat{\bar u}(\cdot,\tilde t)-\bar u(\cdot,\tilde t).
\]
Then for all $\tilde t\in[0,T_{\mathrm{out}}]$,
\begin{equation}\label{eq:state_flux_relation_exact}
\delta u(\cdot,\tilde t)=\tilde t\,\nabla\cdot \delta\psi(\cdot,\tilde t).
\end{equation}
On a periodic domain $\Omega=\mathbb{T}^d$, let $\widehat{\delta\psi}(k,\tilde t)$ be the Fourier coefficient at
$k\in\mathbb{Z}^d$, and define the longitudinal (divergence-relevant) component
\[
\widehat{\delta\psi}_{\parallel}(k,\tilde t):=
\frac{k}{|k|}\Big(\frac{k}{|k|}\cdot \widehat{\delta\psi}(k,\tilde t)\Big),\qquad k\neq 0,
\]
with $\widehat{\delta\psi}_{\parallel}(0,\tilde t):=0$. Then
\begin{equation}\label{eq:l2_equals_weighted_flux}
\|\delta u(\cdot,\tilde t)\|_{L^2(\mathbb{T}^d)}^2
=
\tilde t^2\sum_{k\in\mathbb{Z}^d\setminus\{0\}} |k|^2\,\big|\widehat{\delta\psi}_{\parallel}(k,\tilde t)\big|^2.
\end{equation}
\end{property}

\begin{proof}
Subtract~\eqref{eq:update_rule} from~\eqref{eq:update_rule_hat} to obtain~\eqref{eq:state_flux_relation_exact}.
On $\mathbb{T}^d$, taking Fourier transforms gives
$\widehat{\delta u}(k,\tilde t)=\tilde t\,(ik\cdot \widehat{\delta\psi}(k,\tilde t))$.
Since $k\cdot \widehat{\delta\psi}(k,\tilde t)=|k|\big(\frac{k}{|k|}\cdot \widehat{\delta\psi}(k,\tilde t)\big)$
and $\big|\frac{k}{|k|}\cdot \widehat{\delta\psi}(k,\tilde t)\big|=|\widehat{\delta\psi}_{\parallel}(k,\tilde t)|$,
Parseval's identity yields~\eqref{eq:l2_equals_weighted_flux}.
\end{proof}

Property~\ref{prop:mode_weighting} makes explicit that an $L^2$ mismatch in the \emph{state} corresponds to a
$|k|^2$-weighted mismatch of the \emph{divergence-relevant component} of the effective flux. As a result, high-wavenumber
errors in the learned effective flux induce amplified penalties in state space, which effectively encourages
high-frequency mode matching in operator training. This built-in mode weighting can help counteract the tendency of
neural optimization to prioritize low-frequency components (spectral bias), without introducing an explicit
frequency-weighted loss.

\subsection{Memory length estimation}\label{sec:memory}
A central modeling choice in memory-aware operator learning is the memory length $T_{\mathrm{in}}$, which should be long enough to capture the effective memory induced by coarse-graining, yet short enough to remain data-efficient. Motivated by the Mori--Zwanzig viewpoint, we estimate $T_{\mathrm{in}}$ from the temporal decorrelation of the \emph{effective closure forcing} $r(\cdot,t)$ that represents the influence of unresolved scales on resolved dynamics.

To obtain a robust scalar diagnostic, we summarize $r$ by the instantaneous energy injection rate into the resolved scales,
\begin{equation}\label{eq:inj_rate}
q(t) := \langle r(\cdot,t), \bar u(\cdot,t)\rangle,
\end{equation}
where $\langle a,b\rangle=\int_{\Omega} a(x)b(x)\,dx$ denotes the $L^2$ inner product on the spatial domain $\Omega$. This quantity measures how unresolved processes act on the resolved energy budget and is less sensitive to phase drift than pointwise field correlations.

Given a time series $q(t_n)$ sampled at uniform spacing $\Delta t$, we form the normalized temporal autocorrelation
\begin{equation}\label{eq:autocorr}
\rho(\tau_\ell) \;=\;
\frac{\mathbb{E}\big[(q(t_n)-\bar q)\,(q(t_{n+\ell})-\bar q)\big]}
     {\mathbb{E}\big[(q(t_n)-\bar q)^2\big]},
\qquad \tau_\ell = \ell\,\Delta t,
\end{equation}
where $\bar q$ is the temporal mean and $\mathbb{E}[\cdot]$ denotes an empirical average over time (and, when available, an ensemble average over trajectories). We then define the \emph{integral decorrelation time} by
\begin{equation}\label{eq:int_time}
\tau_{\mathrm{decc}} := \int_{0}^{\infty} \rho(\tau)\,d\tau,
\end{equation}
approximated numerically using the discrete $\rho(\tau_\ell)$. In practice, the integral (\ref{eq:int_time}) must be approximated over a finite lag interval. When the normalized autocorrelation $\rho(\tau)$ is nonoscillatory and decays approximately monotonically, a first-zero truncation, i.e. integration over $[0,\tau_0]$ with $\tau_0:=\inf \{\tau > 0 : \rho(\tau) = 0\}$, provides a conservative estimate of the decorrelation time. By contrast, when $\rho(\tau)$ is strongly oscillatory and exhibits an early zero-crossing followed by substantial positive lobes, the first-zero rule can terminate the integral prematurely. In that case, a non-negative-lobe truncation, which retains the non-negative portions of $\rho(\tau)$, is more appropriate because it preserves the contribution of delayed positive correlation and thus yields a more faithful estimate of the large-scale memory time. Physical mechanisms such as coherent vortical structures or pronounced periodicity may produce this oscillatory correlation pattern, but the truncation criterion is the shape of $\rho(\tau)$ itself.

We set the memory length proportional to the integral time scale,
\begin{equation}\label{eq:Tin_choice}
T_{\mathrm{in}} \in [\tau_{\mathrm{decc}}, 5\tau_{\mathrm{decc}}],
\end{equation}
and choose the specific value within this range by a lightweight grid search. This procedure yields a data-driven guideline for memory-length selection that is consistent with a coarse-graining-induced, finite-memory approximation of non-Markovian effects.

\section{Numerical experiments}
\label{sec:experiments}
We evaluate the proposed causal and locally conservative spatiotemporal operator learning approach on three coarse-grained PDE benchmarks: the one-dimensional viscous Burgers' equation, the one-dimensional Kuramoto--Sivashinsky (KS) equation, and the two-dimensional Navier--Stokes (NS) equations. Viscous Burgers' provides a controlled advective–dissipative setting where prediction fidelity and energy decay can be assessed directly. KS is chaotic, so pointwise errors are only expected to remain bounded and statistical agreement becomes the primary criterion. NS adds the complexity of two-dimensional turbulent dynamics

In all three cases, the learning target is the filtered field obtained by the sharp spectral projector $\mathcal{P}_{k_{\mathrm{cut}}}$:
\begin{equation}\label{eq:burgers_filter}
\bar u(\cdot,t) = \mathcal{P}_{k_{\mathrm{cut}}} u(\cdot,t),
\qquad 
\mathcal{P}_{k_{\mathrm{cut}}}:\ \mathcal{F}[u(\cdot,t)]_k \mapsto \mathbf{1}_{|k|\le k_{\mathrm{cut}}}\,\mathcal{F}[u(\cdot,t)]_k,
\end{equation}
with cutoff $k_{\mathrm{cut}}=12$ for Burgers' and KS, and $k_{\mathrm{cut}}=8$ for NS. This filter removes high-wavenumber content while preserving the low-wavenumber structure that controls the large-scale transport and pattern organization.

For model training, we use a Sobolev-type data-misfit that augments the $L^2(\Omega)$ error on the resolved field with matching of low-order spatial derivatives. Concretely, for a predicted future slab $\widehat{\bar u}_0^{T_{\mathrm{out}}}$ and reference $\bar u_0^{T_{\mathrm{out}}}$, we minimize the sum of
\[
\|\widehat{\bar u}_0^{T_{\mathrm{out}}}-\bar u_0^{T_{\mathrm{out}}}\|^2_{L^2},
\qquad
\|\partial_x\widehat{\bar u}_0^{T_{\mathrm{out}}}-\partial_x\bar u_0^{T_{\mathrm{out}}}\|^2_{L^2},
\qquad
\|\partial_{xx}\widehat{\bar u}_0^{T_{\mathrm{out}}}-\partial_{xx}\bar u_0^{T_{\mathrm{out}}}\|^2_{L^2}.
\]
This derivative-augmented loss is inspired by prior work on learning dissipative chaotic dynamics~\cite{li2022learning}, where incorporating gradient information improves long-horizon stability and reduces spurious high-frequency artifacts. All spatial derivatives appearing in the loss and in the divergence-form update are evaluated spectrally. On the periodic domain, we compute spatial gradients and divergences by FFT: in Fourier space, $\partial_x$ (and more generally $\nabla$) acts as multiplication by $ik$, followed by an inverse FFT to return to physical space. In all experiments, the number of retained Fourier modes in each FNO layer is set equal to the coarse-graining cutoff $k_{\mathrm{cut}}$ (i.e., we retain modes $|k|\le k_{\mathrm{cut}}$ in the spectral convolution). Thus the operator layers do not access spatial frequencies beyond those present in the coarse-grained state. Throughout, we use the same optimizer and learning-rate schedule: Adam~\cite{kingma2015adam} with initial learning rate $10^{-3}$ and a cosine-annealing schedule~\cite{loshchilov2017sgdr} over $500$ epochs.

Throughout this section, we label our model as \texttt{Local conservation}, and study the role of conservative inductive bias by comparing it with two baselines:
\begin{itemize}
  \item \texttt{No conservation}; this model takes the form
  \begin{equation}\label{eq:baseline1}
    \widehat{\bar u}(\cdot,\tilde t) = \bar u(\cdot,0) + \tilde t\,\widehat{\Phi}\!\big(\bar u^0_{-T_{\mathrm{in}}}\big)(\cdot,\tilde t),
  \end{equation}
  where $\widehat{\Phi}$ is a purely data-driven spatiotemporal neural operator, ignoring the conservative inductive bias.
  \item \texttt{Global conservation}; the second baseline model takes the form
  \begin{equation}\label{eq:baseline2}
    \widehat{\bar u}(\cdot,\tilde t) = \bar u(\cdot,0) + \tilde t\,\left(\widehat{\Phi}\!\big(\bar u^0_{-T_{\mathrm{in}}}\big)(\cdot,\tilde t) - \frac{1}{\mathrm{vol}(\Omega)}\int_\Omega\,\widehat{\Phi}\!\big(\bar u^0_{-T_{\mathrm{in}}}\big)(\cdot,\tilde t) \mathrm{d}x\right),
  \end{equation}
  where the output of $\widehat{\Phi}$ is de-meaned, respecting the global conservation property which holds for all three benchmarks:
  $$\int_\Omega\,\widehat{\bar u}(\cdot,\tilde t)\mathrm{d}x = \int_\Omega\,\bar u(\cdot,0)\mathrm{d}x.$$
\end{itemize}
\subsection{Viscous Burgers\ equation}\label{sec:burgers}
\subsubsection{Dataset description}
\begin{figure}[t]
  \centering
  \includegraphics[width=\linewidth]{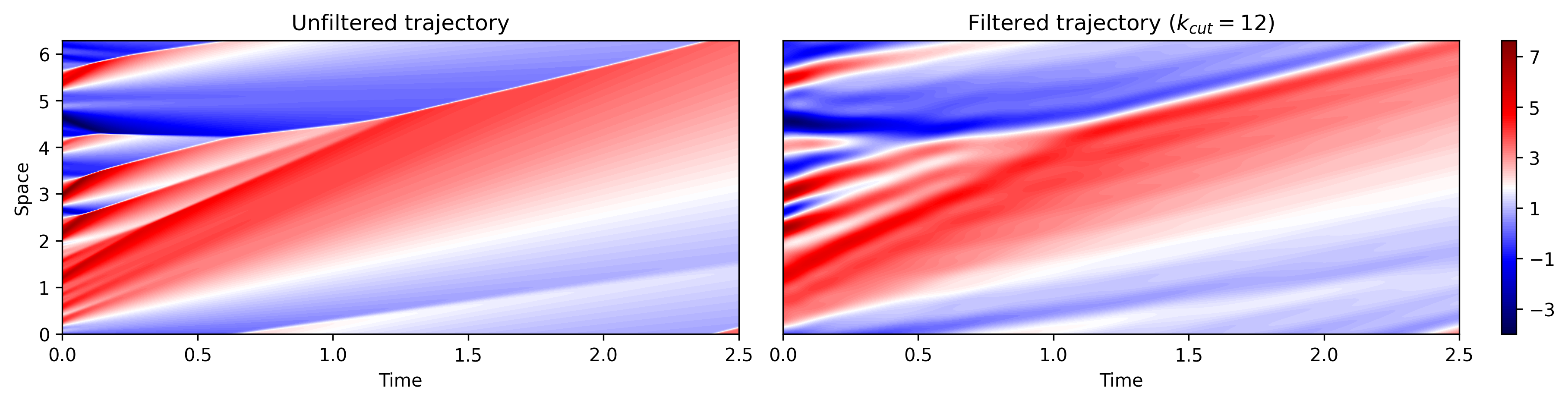}
  \caption{Effect of spectral coarse-graining on the viscous Burgers' dynamics.}
  \label{fig:burgers_raw_filtered}
\end{figure}

We consider the viscous Burgers' equation on the periodic domain $x\in[0,2\pi)$,
\begin{equation}\label{eq:burgers}
u_t + \left(\frac{u^2}{2}-\nu u_x\right)_x = 0, \qquad u(x,0)=u_0(x)
\end{equation}
with viscosity $\nu=10^{-2}$, and generate $110$ reference trajectories on a uniform grid with $L_0=1024$ points and grid size $\Delta x = 2\pi/L_0$. We compute reference solutions using a Fourier pseudo-spectral discretization in space and an exponential time-differencing Runge--Kutta scheme (ETDRK4) in time. The time step size is $\delta=2.5\times 10^{-4}$. The initial conditions are sampled as periodic Gaussian random fields in Fourier space with power spectrum
\begin{equation}\label{eq:grf_spectrum}
S(k)=\sigma^2\bigl(1+(|k|/\tau)^2\bigr)^{-\alpha},
\end{equation}
enforcing conjugate symmetry to obtain real-valued fields. After subtracting the spatial mean, we apply an affine normalization,
\begin{equation}\label{eq:burgers_ic}
u_0(x)= U_0 + 6\,\tilde u_0(x)/\|\tilde u_0\|_{\infty},
\end{equation}
where $U_0=2$ and $\tilde u_0$ denotes the sampled zero-mean field. This yields trajectories with a nontrivial mean flow and $\mathcal{O}(1)$ fluctuations. The resulting dynamics exhibit the canonical advect--steepen--diffuse behavior of viscous Burgers'; see Figure~\ref{fig:burgers_raw_filtered}. The background flow induces a dominant right-going transport so that coherent structures drift across the periodic domain. Superimposed on this drift, the quadratic flux produces rapid gradient steepening that concentrates activity into localized viscous shock layers, while viscosity regularizes these fronts and removes high-wavenumber content. As a result, the resolved low-frequency field $\bar u$ displays propagating, intermittently steepened features with gradual amplitude decay.

The dataset is further downsampled in space by a factor of $4$ and in time by a factor of $10$, leading to a grid size of $L=256$ and a time lag of $\Delta_{\text{train}}=2.5\times 10^{-3}$. We use the first $10$ trajectories for training and the last $100$ for evaluation. 

\subsubsection{Model training and evaluation}
Using the memory length estimation procedures presented in Section \ref{sec:memory}, we find the integral decorrelation time $\tau_{\mathrm{decc}}$ is about $0.2$. We thus set $T_{\mathrm{in}}=0.25$ and $T_{\mathrm{out}}=0.125$. In each training sample, the numbers of input and output snapshots are
\begin{equation}
m = \mathrm{round}(T_{\mathrm{in}}/\Delta_{\text{train}})+1 = 101,
\qquad
s = \mathrm{round}(T_{\mathrm{out}}/\Delta_{\text{train}}) = 50.
\end{equation}
A total of $J=1,000$ training samples are randomly sampled from the first $10$ reference trajectories. We instantiate a spatiotemporal neural operator with two spatiotemporal blocks (depth $2$) and width $64$. To match the resolved low-frequency dynamics, we project model outputs onto the same resolved spectral subspace (i.e., the first $k_{\mathrm{cut}}$ Fourier modes), which is effectively implemented as a nontrainable layer of the neural operator. We train the model for $500$ epochs with a batch size $10$.

Once the training is finished, we evaluate the trained model on the held-out $100$ testing trajectories by recursively applying the learned evolution operator to generate predictions up to time $T=2.5$. To quantify accuracy, we report the time-dependent relative error in state space
\begin{equation}\label{eq:relerr}
\varepsilon(t) = \frac{\|\widehat{\bar u}(\cdot,t)-\bar u(\cdot,t)\|_{L^2}}{\|\bar u(\cdot,t)\|_{L^2}},
\end{equation}
and the mode-wise energy absolute error:
\begin{equation}\label{eq:energyrmse}
\eta(t, k) = \left|\hat E_k(t)-E_k(t)\right|^2,\quad \hat E_k(t) = \frac{1}{2}\,\Big|\mathcal{F}[\widehat{\bar u}(\cdot,t)]_k\Big|^2,\quad E_k(t) = \frac{1}{2}\,\Big|\mathcal{F}[\bar u(\cdot,t)]_k\Big|^2,
\end{equation}
for $k=1,\ldots,k_{\mathrm{cut}}$.
Both metrics are averaged over $100$ testing trajectories.

\begin{figure}[t]
  \centering
  \includegraphics[width=\linewidth]{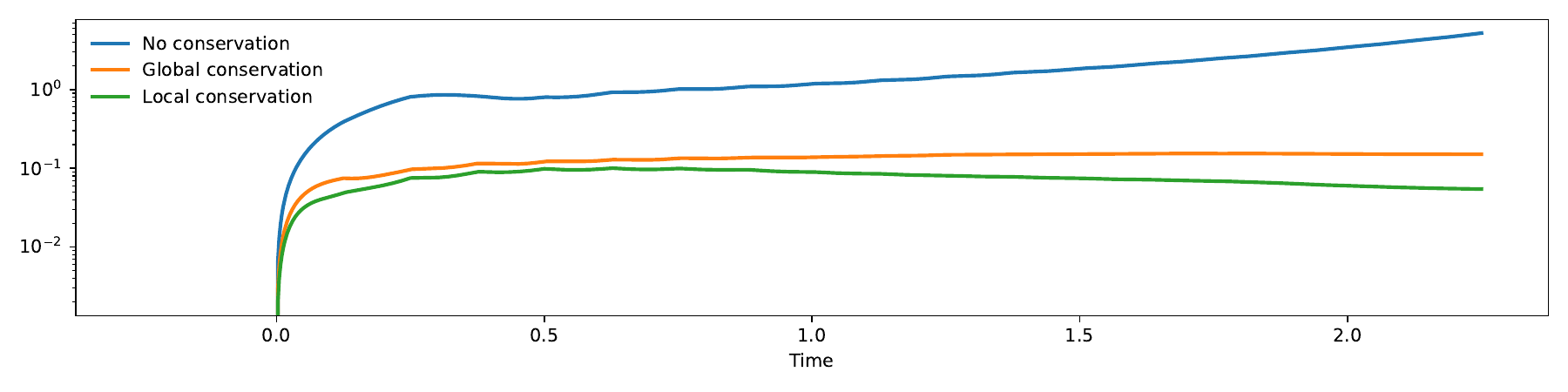}
  \caption{Viscous Burgers': Relative $L^2$ error $\varepsilon(t)$ in state space during the rollout.}
  \label{fig:burgers_relerr}
\end{figure}

\begin{figure}[!htbp]
  \centering
  \includegraphics[width=\linewidth]{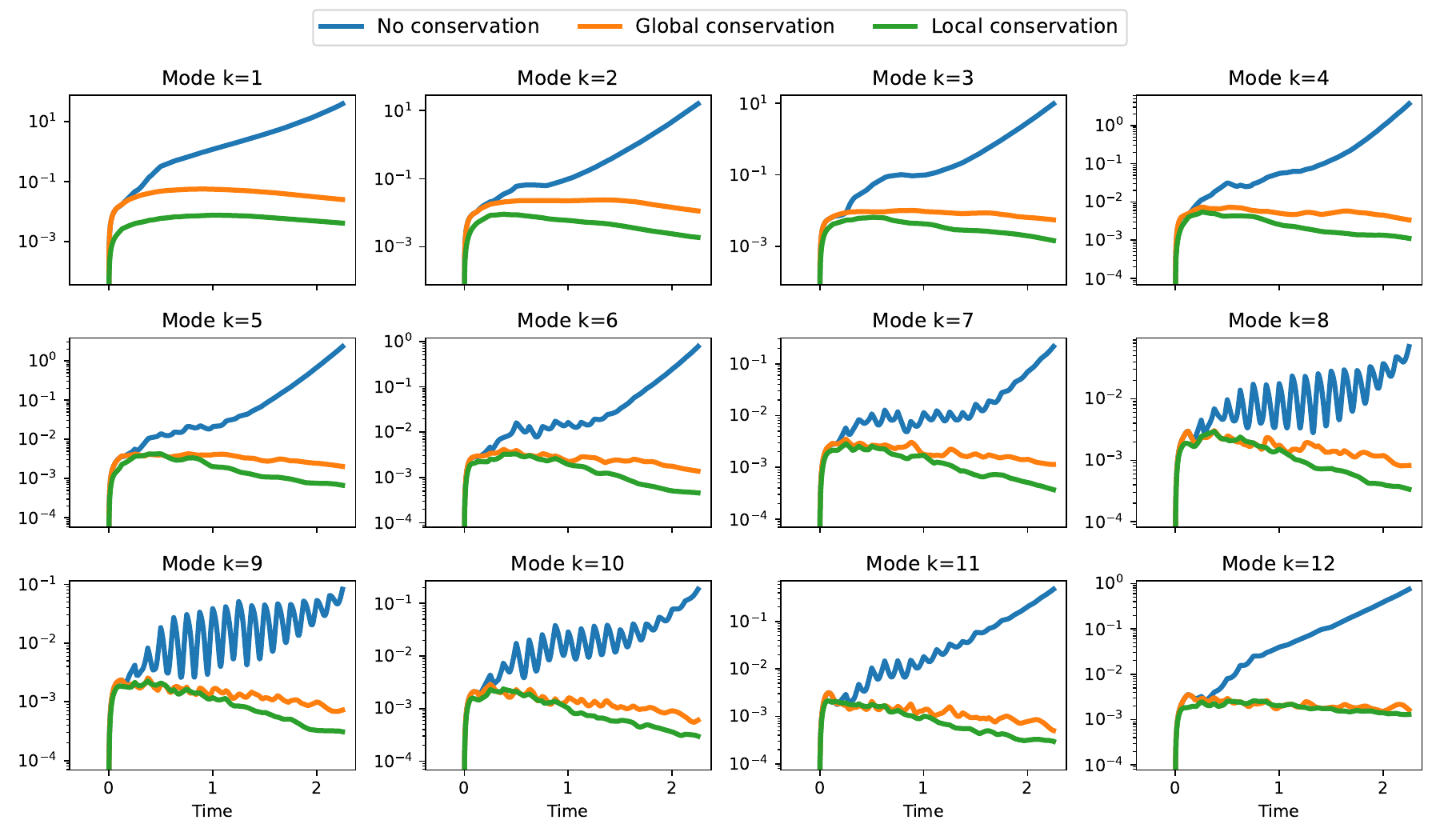}
  \caption{Viscous Burgers': Mode-wise absolute error of resolved energies.}
  \label{fig:burgers_energy_error}
\end{figure}

As in Figure \ref{fig:burgers_relerr}, the relative error of both \texttt{Global conservation} and \texttt{Local conservation} grow slowly with time, remaining small throughout the rollout and saturating rather than diverging like \texttt{No conservation}. Throughout the rollout, \texttt{Local conservation} consistently delivers smaller error than \texttt{Global conservation}. 

This trend is mirrored in the spectral space. In Figure \ref{fig:burgers_energy_error}, we observe that \texttt{Local conservation} achieves the best performance at \emph{every mode} for $k=1,\dots,k_{\mathrm{cut}}$. This indicates that the model does more than fit $\bar u$ in the state space, it also distributes energy correctly across scales.

\begin{figure}[t]
  \centering
  \includegraphics[width=\linewidth]{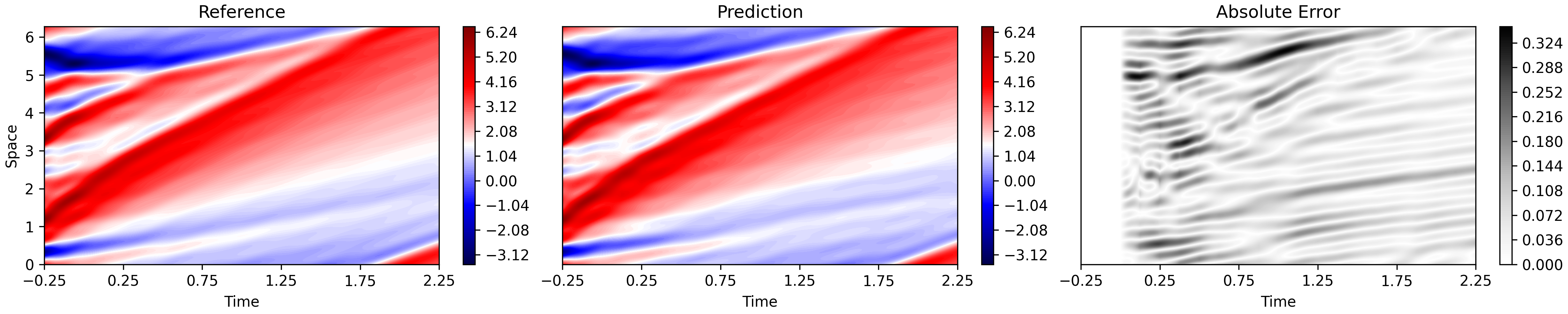}
  \caption{Autoregressive rollout on a held-out filtered Burgers' trajectory.}
  \label{fig:burgers_rollout_contour}
\end{figure}

Beyond ensemble metrics, we present the results of an example from the test set. Figure \ref{fig:burgers_rollout_contour} displays the space--time contours over the full rollout horizon. The model \texttt{Local conservation} reproduces the dominant advective transport and the gradual smoothing associated with viscous dissipation, with only mild phase drift at late times. The absolute error remains localized and does not exhibit the coherent banding patterns typically associated with unstable autoregressive rollouts. The learned operator therefore composes consistently under repeated application

\begin{figure}[!htbp]
  \centering
  \includegraphics[width=\linewidth]{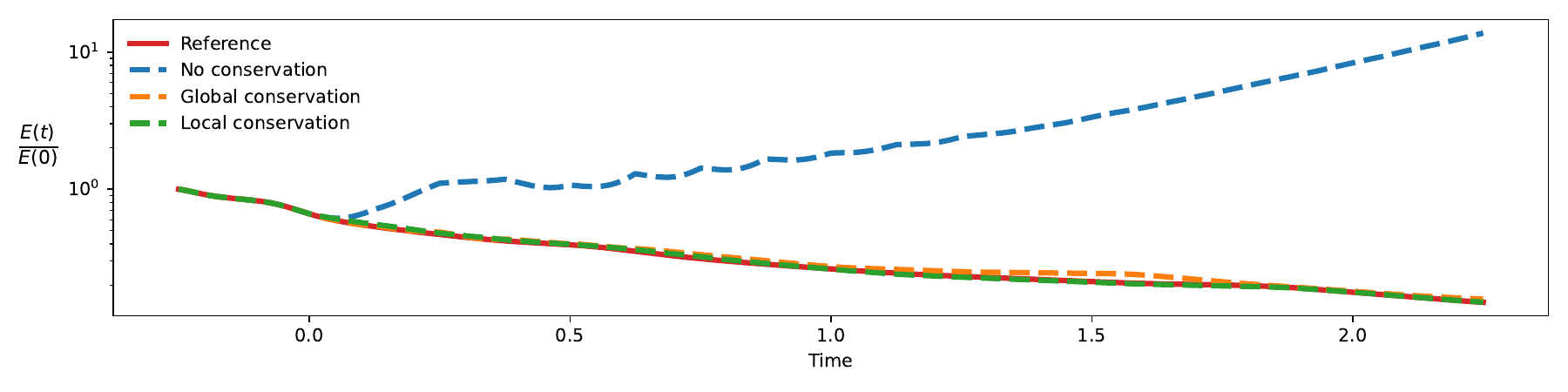}
  \caption{Viscous Burgers': Evolution of resolved total energy $E(t)$, normalized by $E(0)$.}
  \label{fig:burgers_total_energy}
\end{figure}

In Figure \ref{fig:burgers_total_energy}, we show that the model \texttt{Local conservation} accurately tracks the resolved total energy
\begin{equation}\label{eq:energy}
E(t) = \frac{1}{2}\int_{0}^{L} \left(\bar u(x,t)-U_0 \right)^2\,\mathrm{d}x,
\end{equation}
normalized by $E(0)$. The model \texttt{Global conservation} demonstrates visible drift and \texttt{No conservation} fails to capture Burgers' energy decay pattern.

\begin{figure}[!htbp]
  \centering
  \includegraphics[width=\linewidth]{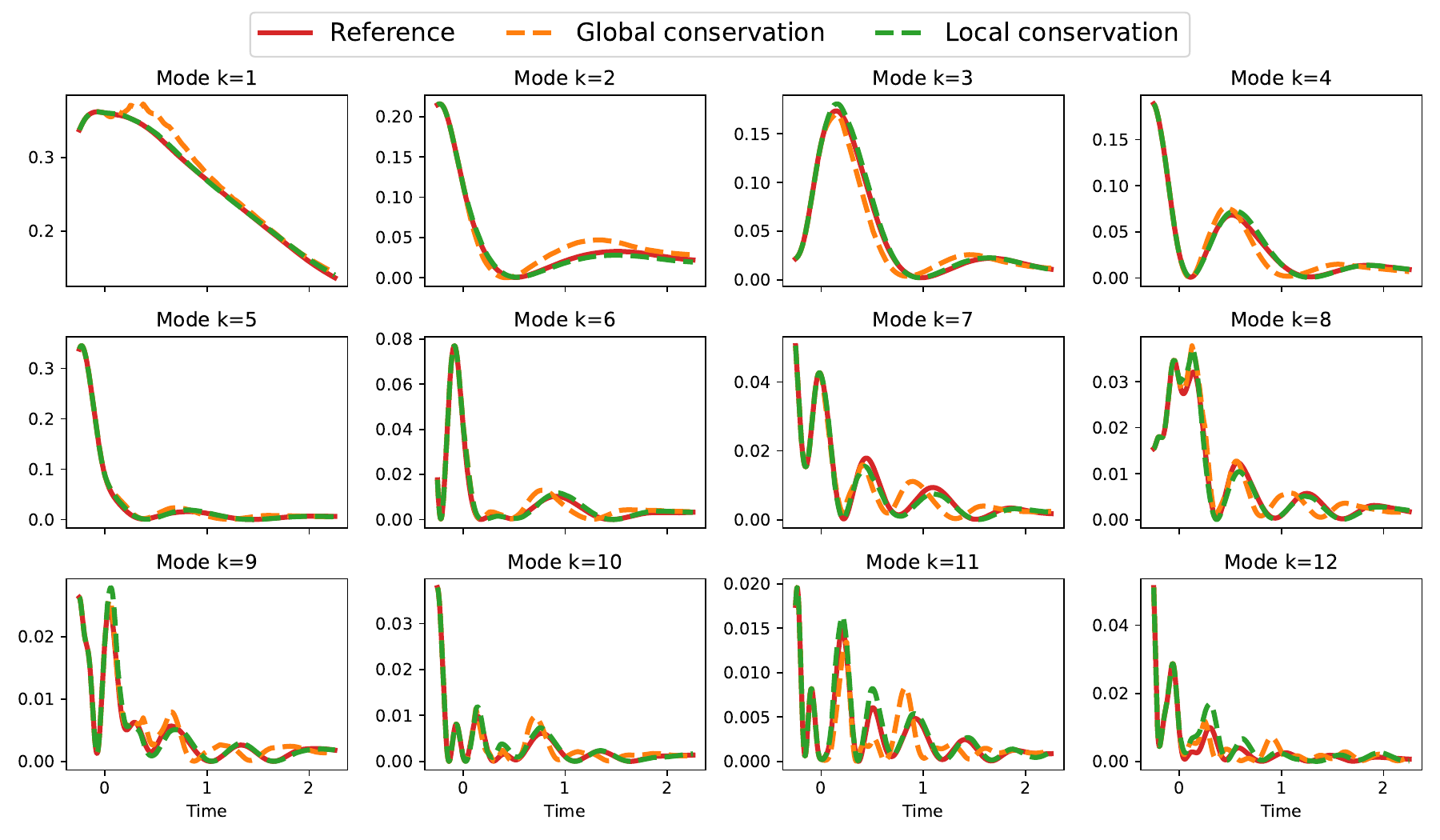}
  \caption{Viscous Burgers': Evolution of mode-wise energy $\frac{1}{2}\,\Big|\mathcal{F}[\bar u(\cdot,t)]_k\Big|^2$ for $1\le k\le k_{\mathrm{cut}}$.}
  \label{fig:burgers_mode_energy}
\end{figure}

For the same example, we display its mode-wise energy evolution in Figure \ref{fig:burgers_mode_energy}. The energy drift of \texttt{Global conservation} becomes more evident, while our proposed approach \texttt{Local conservation} still matches across scales, even at the cutoff mode. 

These results indicate that the locally conservative spatiotemporal operator remains stable and accurate over long rollouts of filtered Burgers' dynamics.

\subsection{Kuramoto--Sivashinsky equation}\label{sec:ks}

\subsubsection{Dataset description}
\begin{figure}[t]
  \centering
  \includegraphics[width=\linewidth]{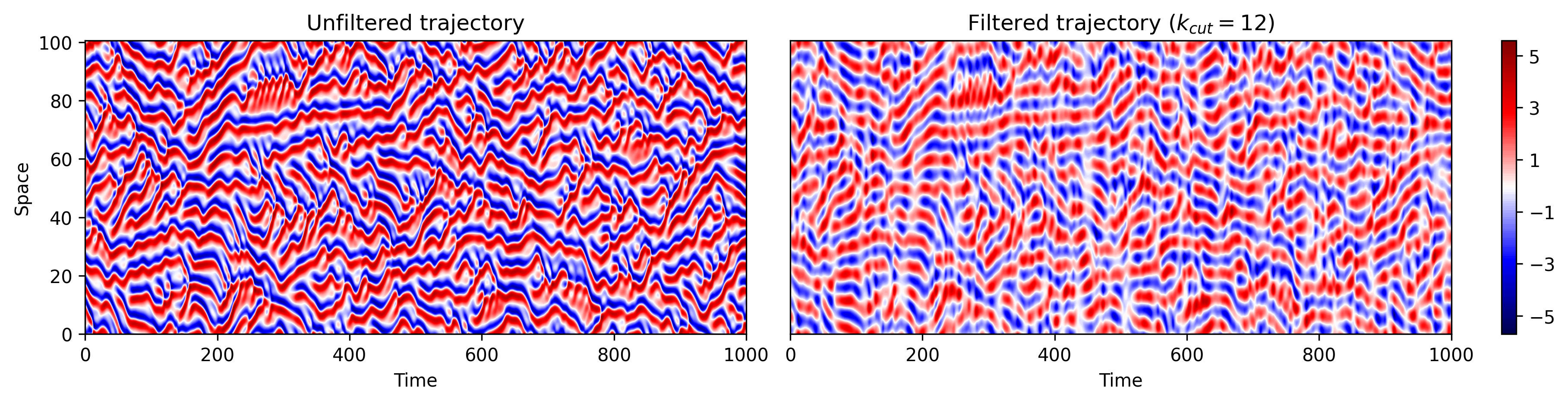}
  \caption{Effect of spectral coarse-graining on the KS dynamics.}
  \label{fig:ks_raw_filtered}
\end{figure}

We consider the Kuramoto--Sivashinsky (KS) equation in conservative form on the periodic domain
$\Omega=[0,32\pi)$:
\begin{equation}\label{eq:ks}
u_t + \partial_x\!\left(\frac{u^2}{2}+u_x+u_{xxx}\right)=0,\qquad x\in[0,32\pi).
\end{equation}
In contrast to the filtered Burgers setting, the KS dynamics in this regime are chaotic and numerically mixing, so long-time statistics are effectively independent of the initial condition. We therefore generate a single long reference trajectory using a Fourier pseudo-spectral discretization with $L=256$ equispaced grid points and ETDRK4 time integration with step size $\delta=10^{-3}$. The initial condition is
\begin{equation}
u(x,0)=\cos(x)\bigl(1+\sin(x)\bigr).
\end{equation}
We integrate until the statistically stationary regime is reached and discard an initial transient. The remaining trajectory is subsampled at a coarse increment $\Delta_{\mathrm{train}}=0.1$, yielding $110{,}000$ snapshots (total duration $11{,}000$). We use the first $10{,}000$ time units for training and reserve the subsequent $1{,}000$ time units for evaluation. An example of the spatiotemporal evolution of the fully resolved and filtered trajectories is shown in Figure~\ref{fig:ks_raw_filtered}, where the effect of spectral coarse-graining removes fine-scale filamentary features while retaining the large-scale organization.

\subsubsection{Model training and evaluation}
Using the memory-length estimation procedure in Section \ref{sec:memory}, we obtain an integral decorrelation time of approximately $\tau_{\mathrm{decc}}\approx 2.9$. We set $T_{\mathrm{in}}=5$ and $T_{\mathrm{out}}=1.5$. Each training pair consists of a history slab with $m=T_{\mathrm{in}}/\Delta_{\mathrm{train}}+1=51$ snapshots and a future slab with $s=T_{\mathrm{out}}/\Delta_{\mathrm{train}}=15$ snapshots. We extract $4{,}500$ training samples by uniformly sampling such space--time windows from the training segment.

We instantiate the spatiotemporal neural operator with depth $3$ and hidden width $64$ and train with batch size $25$. We evaluate by autoregressive rollout on the held-out segment over $T=1000$ time units, which requires composing the learned operator $T/T_{\mathrm{out}}\approx 667$ times. Because KS is chaotic, pointwise trajectory agreement is not expected beyond a short horizon; we therefore assess (i) rollout stability and (ii) statistical fidelity of the predicted resolved dynamics.

\begin{figure}[!htbp]
  \centering
  \includegraphics[width=\textwidth]{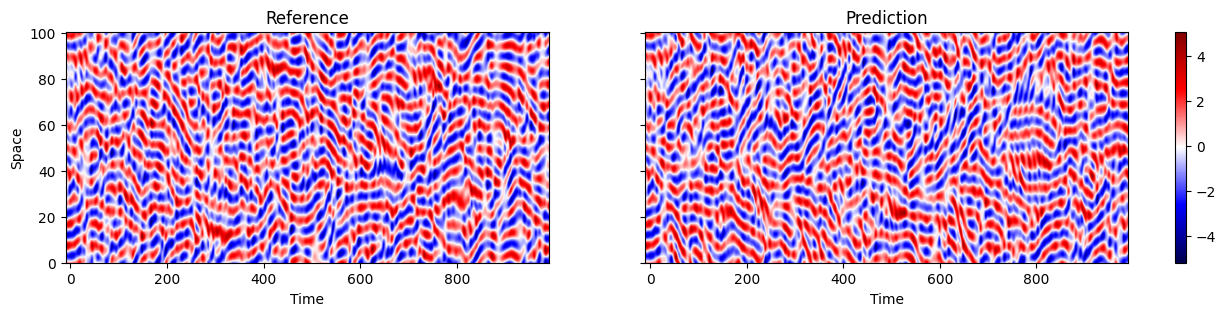}
  \caption{Autoregressive rollout on the held-out filtered KS segment: reference (left) and prediction (right).}
  \label{fig:ks_rollout_contour}
\end{figure}

\begin{figure}[!htbp]
  \centering
  \includegraphics[width=\linewidth]{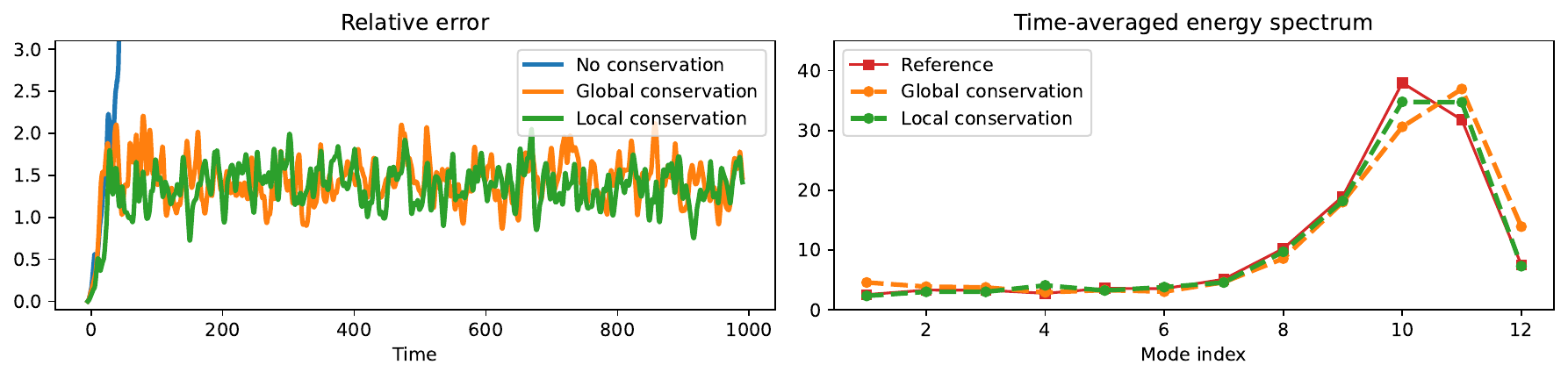}
  \caption{Kuramoto--Sivashinsky: (left) relative $L^2$ rollout error $\varepsilon(t)$; (right) time-averaged resolved energy spectrum $S(k)$.}
  \label{fig:ks_relerr_energy}
\end{figure}

Figure~\ref{fig:ks_rollout_contour} compares the space--time evolution on the test segment. The \texttt{Local conservation} model remains in the correct spatiotemporal regime throughout the rollout, without collapse to a trivial state or numerical blow-up. The relative field error $\varepsilon(t)$ is shown in Figure~\ref{fig:ks_relerr_energy} (left). The baseline \texttt{No conservation} becomes unstable early, while both \texttt{Global conservation} and \texttt{Local conservation} saturate at an $\mathcal{O}(1)$ level, consistent with exponential separation of trajectories on the KS attractor rather than systematic numerical drift.

We report three complementary long-time statistics computed on the $1{,}000$-unit test segment (i.e., $10{,}000$ snapshots), and on the corresponding predicted rollout.

\emph{(i) Time-averaged energy spectrum.}
Let $\mathcal{F}[\bar u(\cdot,t)]_k$ denote the spatial Fourier coefficients (with $k\ge 1$). We define the time-averaged resolved spectrum by
\begin{equation}\label{eq:ks_spectrum}
S(k):=\mathbb{E}_t\left[\Big|\mathcal{F}\big[\bar u(\cdot,t)\big]_k\Big|^2\right],
\end{equation}
where $\mathbb{E}_t[\cdot]$ denotes empirical averaging over the test interval. We summarize spectral mismatch by the relative $\ell_2$ error over by taking $S$ as a $k_{\mathrm{cut}}-$dimensional vector:
\begin{equation}\label{eq:ks_spec_err}
  \varepsilon_S:=\frac{\|S_{\mathrm{pred}}-S_{\mathrm{ref}}\|_{\ell_2}}{\|S_{\mathrm{ref}}\|_{\ell_2}}
\end{equation}
Figure~\ref{fig:ks_relerr_energy} (right) shows that \texttt{Local conservation} matches the reference spectrum more accurately than \texttt{Global conservation}, including around the peak-energy mode and the high-wavenumber tail. This is consistent with the flux-form parameterization, which biases the learning toward improved high-mode fidelity (cf.\ Section \ref{sec:methods_increment}).

\emph{(ii) Time-averaged spatial autocorrelation and integral length scale.}
Define the snapshot-centered field $u'(x,t):=\bar u(x,t)-\frac{1}{|\Omega|}\int_\Omega \bar u(y,t)\,dy$ and the (periodic) normalized spatial autocorrelation at distance $r$:
\begin{equation}\label{eq:ks_spatial_acf}
\rho_x(r;t):=\frac{\langle u'(\cdot,t),u'(\cdot+r,t)\rangle}{\langle u'(\cdot,t),u'(\cdot,t)\rangle},
\qquad
\langle a,b\rangle:=\int_\Omega a(x)b(x)\,dx.
\end{equation}
We then time-average the per-snapshot normalized correlation,
\begin{equation}\label{eq:ks_spatial_acf_timeavg}
\rho_x(r):=\mathbb{E}_t\bigl[\rho_x(r;t)\bigr],\qquad r\in[0,|\Omega|/2],
\end{equation}
and define the integral length scale by integrating up to the first zero crossing:
\begin{equation}\label{eq:ks_int_length}
\ell_{\mathrm{int}}:=\int_0^{r_0}\rho_x(r)\,dr,
\qquad
r_0:=\inf\{r>0:\rho_x(r)=0\}.
\end{equation}

\emph{(iii) Temporal autocorrelation of resolved total energy and integral time scale.}
Define the resolved total energy
\begin{equation}\label{eq:ks_energy}
E(t):=\frac{1}{2}\|\bar u(\cdot,t)\|_{L^2(\Omega)}^2=\frac{1}{2}\int_\Omega \bar u(x,t)^2\,dx,
\qquad
E'(t):=E(t)-\mathbb{E}_t[E(t)].
\end{equation}

\begin{figure}[t]
  \centering
  \includegraphics[width=\linewidth]{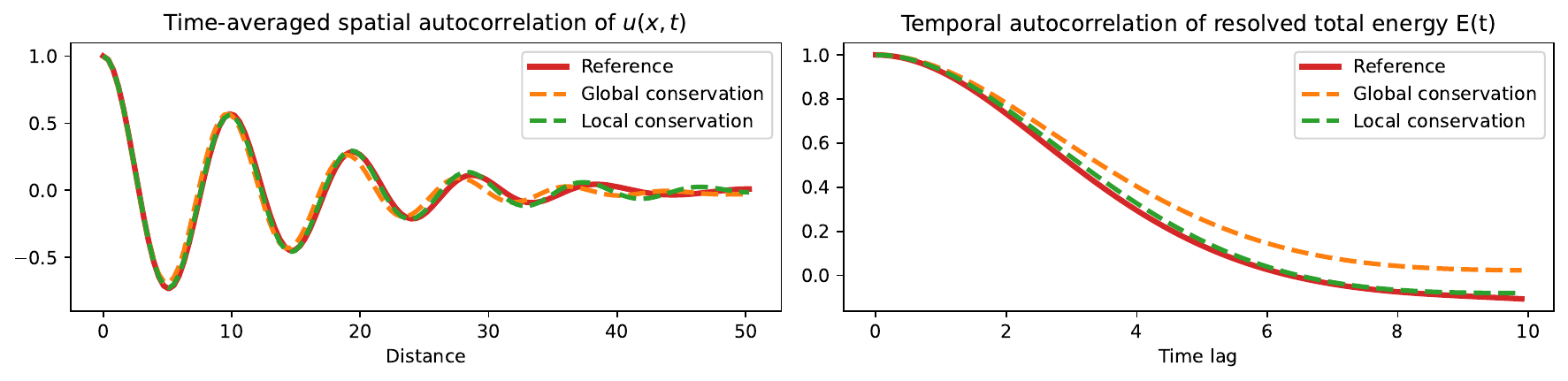}
  \caption{Two-point statistics of the filtered KS dynamics: time-averaged spatial autocorrelation (left) and temporal autocorrelation of resolved energy fluctuation (right).}
  \label{fig:ks_autocorr}
\end{figure}

\begin{table}[!htbp]
\footnotesize
\caption{Quantitative diagnostics on filtered KS dynamics. The integral length scale $\ell_{\mathrm{int}}$ is defined in \eqref{eq:ks_int_length}, the energy integral time scale $\tau_E$ in \eqref{eq:ks_int_time}, and the energy spectrum error $\varepsilon_S$ in \eqref{eq:ks_spec_err}.}
\label{tab:ks_metrics}
\begin{center}
\setlength{\tabcolsep}{4pt} 
\renewcommand{\arraystretch}{1.2} 

  \begin{tabular}{|c|c|c|c|} \hline
   & \bfseries \thead{Spatial integral\\length scale $\ell_{\mathrm{int}}$} 
   & \bfseries \thead{Energy's integral\\ time scale $\tau_E$} 
   & \bfseries \thead{Energy spectrum's \\relative error $\varepsilon_S (\downarrow)$} \\ \hline
   
    \texttt{Global conservation} & $1.717$ & $3.765$ & $0.2059$ \\ 
    \texttt{Local conservation} & $\mathbf{1.728}$ & $\mathbf{3.238}$ & $\mathbf{0.08465}$\\  \hline
    
    \texttt{Reference} & $1.729$ & $3.112$ & \diagbox[dir=SW, width=10.5em, height=2em]{}{} \\ \hline
  \end{tabular}
\end{center}
\end{table}

\noindent The normalized temporal autocorrelation of energy fluctuations is
\begin{equation}\label{eq:ks_energy_acf}
\rho_E(\tau):=\frac{\mathbb{E}_t\bigl[E'(t)\,E'(t+\tau)\bigr]}{\mathbb{E}_t\bigl[(E'(t))^2\bigr]},
\qquad \tau\ge 0,
\end{equation}
and the corresponding integral time scale is
\begin{equation}\label{eq:ks_int_time}
\tau_E:=\int_0^{\tau_0}\rho_E(\tau)\,d\tau,
\qquad
\tau_0:=\inf\{\tau>0:\rho_E(\tau)=0\}.
\end{equation}

Figure~\ref{fig:ks_autocorr} shows that \texttt{Local conservation} reproduces more accurately the dominant spatial correlation structure and the decay of the energy autocorrelation. Table~\ref{tab:ks_metrics} summarizes $\ell_{\mathrm{int}}$, $\tau_E$, and $\varepsilon_S$. Overall, \texttt{Local conservation} achieves the best spectral fidelity while maintaining stable long-horizon rollouts; the proposed flux-form, memory-aware operator thus recovers the correct invariant statistics of filtered KS dynamics.
\subsection{Incompressible Navier--Stokes equation}\label{sec:ns}

\subsubsection{Dataset description}
We consider the two-dimensional incompressible Navier--Stokes (NS) equation on the periodic domain
$\Omega=[0,2\pi)\times [0,2\pi)$
\begin{equation}\label{eq:ns}
\omega_t + \nabla\cdot\left(\omega \mathbf{v}-\nu\nabla\omega-\mathbf{f}\right)=0,
\end{equation}
with viscosity $\nu=0.002$, and an external forcing term $\mathbf{f}=(0,-\sin(4y))^\top$, leading to strongly nonlinear, chaotic, forced-dissipative vortical dynamics. Due to the periodic boundary condition and the zero spatial mean of the forcing curl, the spatial average of vorticity is preserved:
\begin{equation}\label{eq:ns_conservation}
  \frac{\mathrm{d}}{\mathrm{d}t}\int \omega \,\mathrm{d}x\mathrm{d}y = 0.
\end{equation}

We numerically generate $11$ long reference trajectories with different initial conditions sampled from Gaussian random fields. The reference solutions are computed by a pseudo-spectral method in space and a Crank--Nicolson scheme in time with time step size $\delta=10^{-4}$. The solution data is stored on a $128\times 128$ Cartesian grid with temporal sampling interval $\Delta_{\mathrm{train}}=0.1$. Each trajectory covers a total period of $T=900$. To construct a coarse-grained prediction task, we first apply a spectral cutoff $k_{\mathrm{cut}}\le 8$ and then downsample the filtered snapshots onto a $32\times 32$ grid. The first $10$ trajectories are used for training and the last one is reserved for evaluation. We also compare against a classical large-eddy simulation (LES) baseline using a Smagorinsky subgrid closure evolved directly on the same coarse grid.

\subsubsection{Model training and evaluation}
We adopt the same training protocol as in the previous section. The estimated decorrelation time is about $\tau_{\mathrm{decc}}=12.5$, and the memory length and prediction window are respectively set to
$T_{\mathrm{in}}=10$ and $T_{\mathrm{out}}=1$ to reduce computational cost and running memory consumption. Each training sample therefore consists of a history slab with
$m=T_{\mathrm{in}}/\Delta_{\mathrm{train}}+1=101$ snapshots and a target slab with
$s=T_{\mathrm{out}}/\Delta_{\mathrm{train}}=10$ snapshots. We use a batch size of $50$ and train for $100$ epochs, with validation-loss monitoring to mitigate overfitting. In total, $20{,}000$ training samples are drawn from the first $10$ trajectories. For all three variants, we use the same spatiotemporal neural operator architecture with depth $5$ and hidden width $48$. The trained models are evaluated on the 11th trajectory by autoregressive rollout, requiring $890$ recursive forward passes.

\begin{figure}[t]
  \centering
  \includegraphics[width=\textwidth]{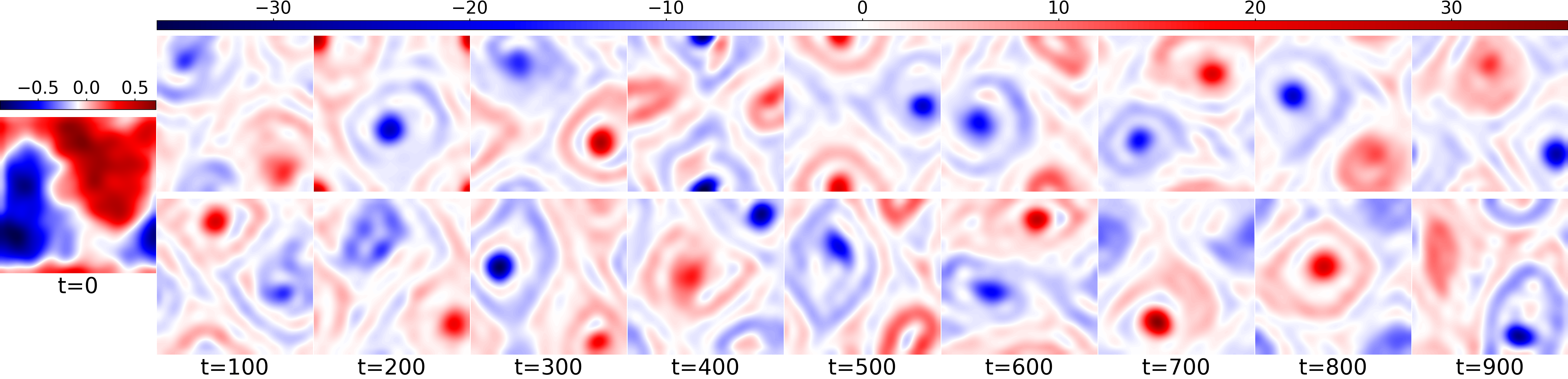}
  \caption{Autoregressive rollout on the held-out filtered NS trajectory.}
  \label{fig:ns_rollout}
\end{figure}

Figure~\ref{fig:ns_rollout} shows representative snapshots from the held-out autoregressive rollout. Starting from the same initial condition (left panel, $t=0$), the prediction remains stable over the entire horizon and generates physically plausible vortical fields up to $t=900$. In particular, the predicted fields retain the qualitative morphology of the forced-dissipative regime, including coherent vortex cores and surrounding filamentary structures, with amplitudes comparable to those of the reference solution. Pointwise agreement deteriorates over long horizons due to sensitivity to initial perturbations, and the locations/phases of individual vortices gradually deviate from the reference trajectory. This loss of phase synchronization is not, by itself, indicative of model failure in the present setting.

\begin{figure}[t]
  \centering
  \includegraphics[width=\textwidth]{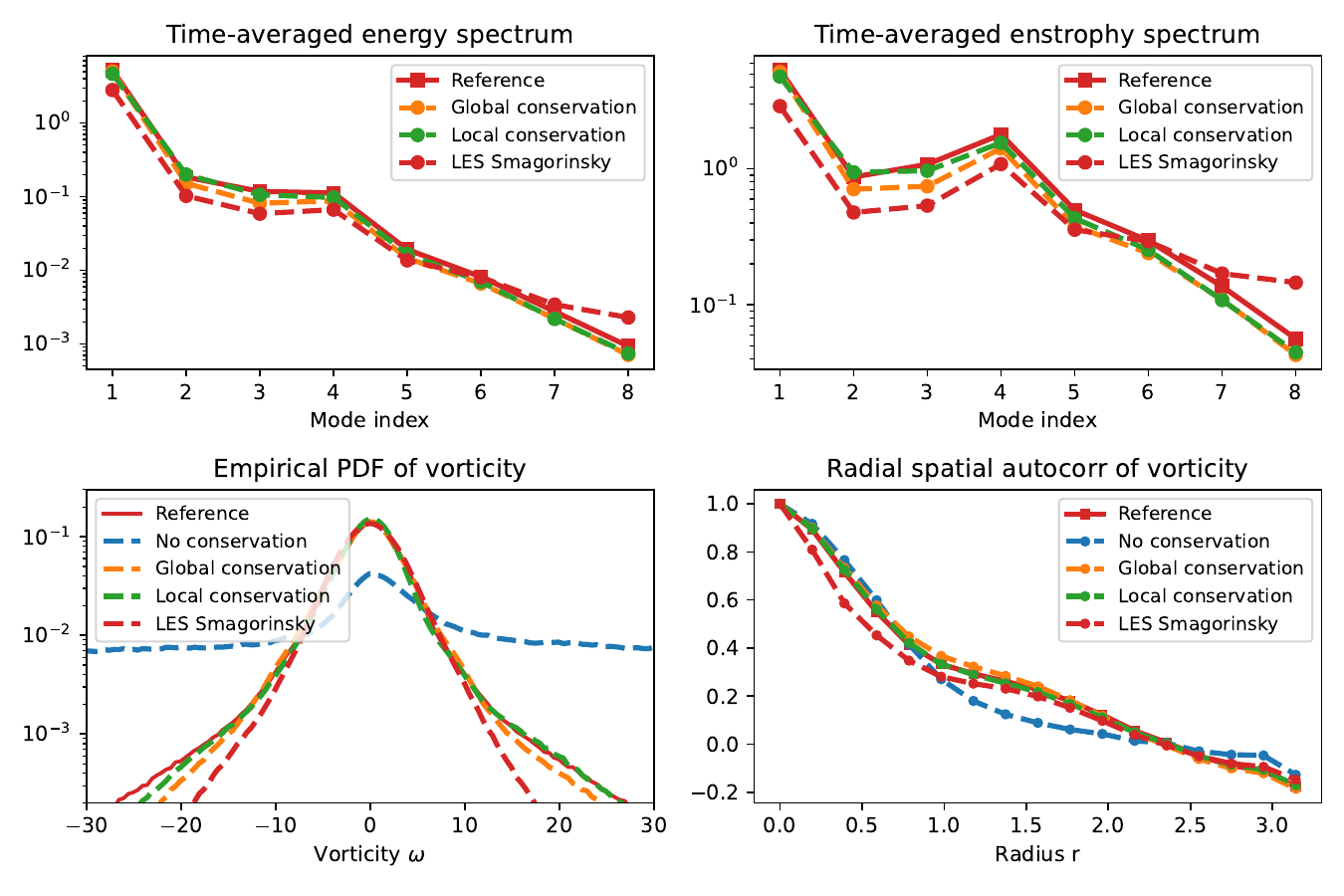}
  \caption{Long-time statistical comparison on the held-out filtered NS trajectory. The learned locally conservative model provides the closest agreement with reference across spectra, vorticity PDF, and spatial autocorrelation, while the LES Smagorinsky baseline remains more dissipative and less accurate at the resolved scales.}
  \label{fig:ns_stats}
\end{figure}

To assess whether the learned surrogate reproduces the effective coarse-grained dynamics, we compare long-time statistics computed from the autoregressive rollout against those of the reference and the LES Smagorinsky baseline. Figure~\ref{fig:ns_stats} reports the time-averaged energy spectrum, the time-averaged enstrophy spectrum, the empirical vorticity PDF, and the radial spatial autocorrelation of vorticity.

The \texttt{No conservation} model fails qualitatively: it produces a distorted stationary distribution, a severely biased spectrum, and an incorrect spatial correlation profile. Enforcing conservation substantially improves all diagnostics. The \texttt{Global conservation} model already restores the correct statistical regime to a large extent, but the \texttt{Local conservation} model is consistently the most accurate learned surrogate across all reported observables.

The LES Smagorinsky baseline performs noticeably better than the unconstrained model and yields physically reasonable coarse dynamics, confirming that the benchmark is nontrivial and that classical subgrid closure captures part of the missing small-scale effect. However, it remains systematically less accurate than the locally conservative model. In the energy spectrum, the LES result underestimates the low-to-intermediate modes, indicating excessive damping of resolved structures. In the enstrophy spectrum, it also shows a visible bias across the resolved range, especially near the higher modes. In the vorticity PDF, the LES profile has thinner tails than the reference, reflecting an over-smoothed stationary state. In the radial spatial autocorrelation, LES decays too rapidly at short distances, showing that it does not preserve spatial coherence accurately.

By contrast, the \texttt{Local conservation} model recovers the low-wavenumber energy content more faithfully than both \texttt{Global conservation} and LES, better reproduces the resolved enstrophy distribution, and yields the best agreement in both the PDF tails and the correlation profile.

This comparison is informative from the closure-modeling viewpoint. The Smagorinsky model represents unresolved dynamics primarily through an eddy-viscosity mechanism, i.e.\ through additional local dissipation. Our locally conservative operator, in contrast, learns a coarse evolution law directly from data while enforcing the correct local balance structure. The improved agreement with reference suggests that the unresolved feedback in this problem is not purely dissipative and cannot be represented optimally by a simple viscosity ansatz. Instead, the learned locally conservative closure appears able to encode more faithful effective transport at the coarse level while still respecting the underlying conservation principle.

\begin{figure}[h!]
  \centering
  \includegraphics[width=0.9\textwidth]{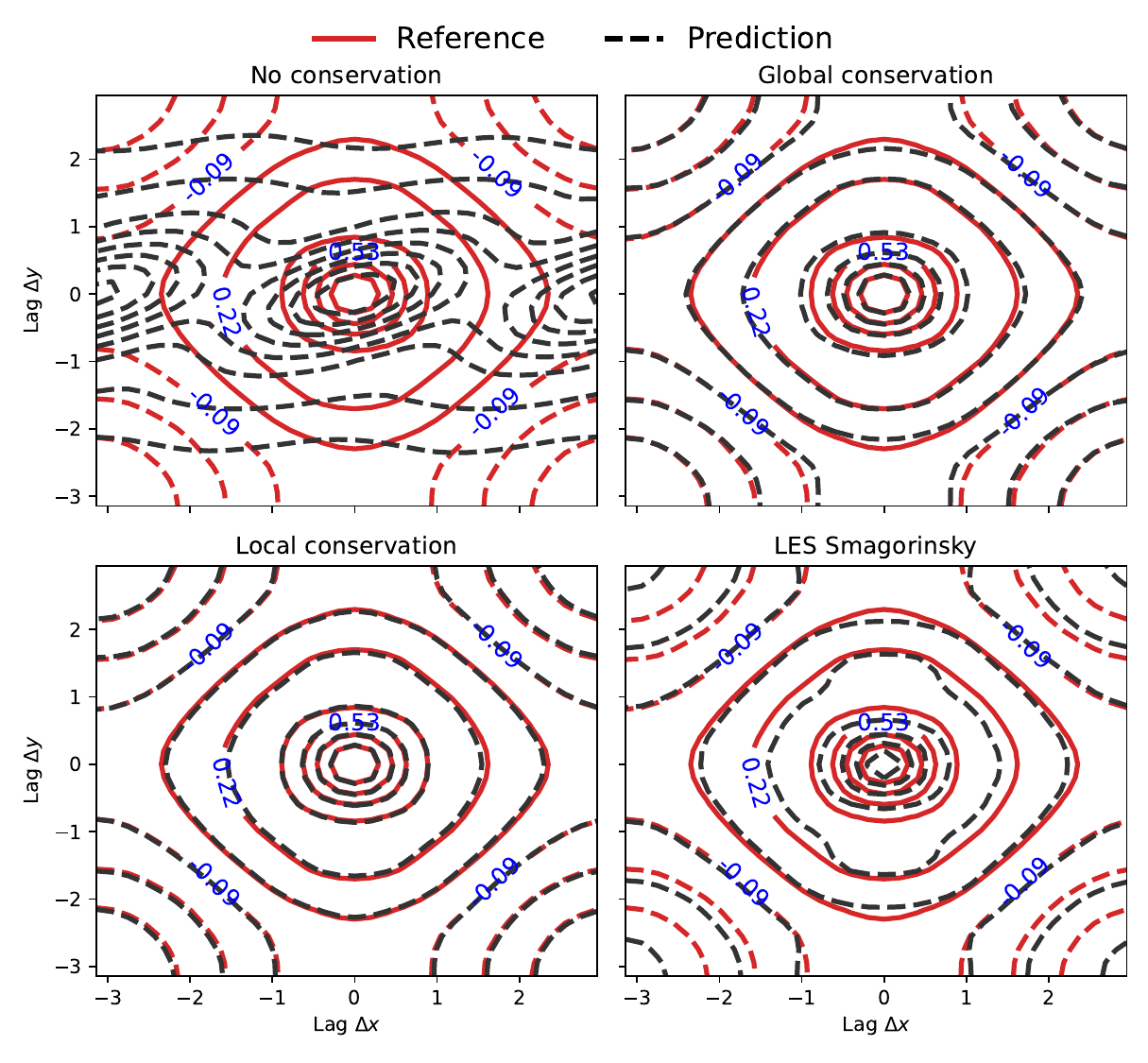}
  \caption{Two-dimensional spatial autocorrelation contours of vorticity. The locally conservative model reproduces the reference contour geometry more accurately than both the globally conservative model and the LES Smagorinsky baseline.}
  \label{fig:ns_corr2d}
\end{figure}

To further examine spatial structure beyond isotropic averaging, Figure~\ref{fig:ns_corr2d} compares the two-dimensional vorticity spatial autocorrelation contours, whose radial averages produce the isotropic profiles shown in Figure~\ref{fig:ns_stats} (bottom-right panel). The differences among the models are more pronounced in this diagnostic. The \texttt{No conservation} model exhibits clear geometric distortion of the correlation field. The LES Smagorinsky baseline captures the overall shape of the contours but still shows visible mismatch in both the core region and the outer levels, again consistent with an overly dissipative closure. The \texttt{Global conservation} model is substantially better aligned with the reference, while the \texttt{Local conservation} model yields contours that nearly overlap the reference across both inner and outer levels. In particular, its agreement at moderate and large lags is the best among all compared methods.

Taken together, these results show that conservation enforcement is essential for learning a faithful coarse-grained NS evolution operator. They further show that local conservation outperforms the global integral constraint and, on this benchmark, is more accurate than a standard physics-based LES closure.

\subsection{Sensitivity to the memory length $T_{\mathrm{in}}$}\label{sec:Tin_sensitivity}
We examine the sensitivity of the learned windowed operator to the choice of memory length $T_{\mathrm{in}}$, while keeping $T_{\mathrm{out}}$ and all other training configurations fixed. The guiding principle from Section~\ref{sec:memory} is to select $T_{\mathrm{in}}$ on the order of a few integral decorrelation times, but the optimal choice is problem dependent.

For filtered Burgers', we report the time-dependent relative $L^2$ rollout error and further average it over the evaluation interval,
\begin{equation}\label{eq:burgers_avg_err}
\varepsilon := \mathbb{E}_t\left[\varepsilon_t\right],
\end{equation}
with $\varepsilon_t$ defined in \eqref{eq:relerr}. For KS, since long-horizon pointwise errors are not informative, we compute time-averaged resolved energy spectra mismatch on the resolved band $k=1,\dots,k_{\mathrm{cut}}$ (excluding $k=0$), as defined in \eqref{eq:ks_spec_err}.

For Burgers, the estimated decorrelation time is $\tau_{\mathrm{decc}}\approx 0.2$. Using $T_{\mathrm{in}}<\tau_{\mathrm{decc}}$ leads to a marked degradation, while $T_{\mathrm{in}}$ of a few $\tau_{\mathrm{decc}}$ yields consistently small errors. For KS, $\tau_{\mathrm{decc}}\approx 2.9$; the best spectrum accuracy is achieved for $T_{\mathrm{in}}$ on the order of $1$--$2$ integral times, with larger windows providing diminishing returns and, in this experiment, slightly worse spectrum matching.

\begin{table}[t]
\centering
\caption{Sensitivity of performance to the memory length $T_{\mathrm{in}}$ (all other settings fixed). Left: viscous Burgers (time-averaged relative $L^2$ error in state space). Right: KS (relative $\ell_2$ error of the time-averaged resolved energy spectrum treated as a 12-dimensional vector over $k=1,\dots,12$).}
\label{tab:Tin_sensitivity}
\begin{minipage}[t]{0.48\textwidth}
\centering
\small
\begin{tabular}{|c|c|c|}
\hline
$T_{\mathrm{in}}$ & $T_{\mathrm{in}}/\tau_{\mathrm{decc}}$ & $\varepsilon (\downarrow)$ \\
\hline
$0.125$ & $0.63$ & $0.3823$ \\
$0.25$  & $1.25$ & $0.07483$ \\
$0.5$   & $2.50$ & $\mathbf{0.06196}$ \\
$1.0$   & $5.00$ & $0.06682$ \\
\hline
\end{tabular}
\end{minipage}\hfill
\begin{minipage}[t]{0.48\textwidth}
\centering
\small
\begin{tabular}{|c|c|c|}
\hline
$T_{\mathrm{in}}$ & $T_{\mathrm{in}}/\tau_{\mathrm{decc}}$ & $\varepsilon_S (\downarrow)$ \\
\hline
$1.5$  & $0.52$ & $0.1582$ \\
$5.0$  & $1.72$ & $\mathbf{0.08465}$ \\
$10.0$ & $3.45$ & $0.1023$ \\
$15.0$ & $5.17$ & $0.1181$ \\
\hline
\end{tabular}
\end{minipage}
\end{table}

\subsection{Effect of temporal history mixing}\label{sec:fno1d2d}
We next isolate the role of explicitly modeling temporal structure in the resolved history. Starting from the \texttt{Local Conservation} architecture, we replace the causal temporal-kernel modules by FNO-style history encodings that pass past snapshots to the network as concatenated channels, but do not apply a dedicated integral operator over history lags. This comparison tests whether the ordered temporal structure of the resolved history improves coarse-grained prediction beyond what can be obtained by stacking or time-tiling past snapshots. Let $\bar u(\cdot,t)\in \mathbb{R}^{d_u}$ denote the resolved state field on $\Omega$, and let $m$ be the number of snapshots contained in the memory window $T_{\mathrm{in}}$. We consider two baseline encodings:
\begin{itemize}
    \item \texttt{FNO1D}: The past $m$ snapshots are concatenated along the channel dimension, producing an $\mathbb{R}^{m d_u}$-valued function on $\Omega$. A spatial FNO is then trained as a one-step predictor for the resolved flow map
    \[
    (\bar{u}_{n-m+1},\ldots,\bar{u}_{n}) \ \mapsto\ \bar{u}_{n+1},
    \]
    where $\bar{u}_{k}=\bar{u}(\cdot,t_k)$ denotes the resolved state field at time $t_k$. In this baseline, temporal information enters only through the ordering of the stacked input channels; the network does not contain an explicit temporal operator acting on history lags.

    \item \texttt{FNO2D}: We adopt the time-tiling construction used in FNO-based sequence prediction~\cite{li2021fourier}. The history window on $[-T_{\mathrm{in}},0]$ is lifted to a space--time input on $\Omega\times[0,T_{\mathrm{out}}]$ by defining a time-constant field
    \[
        \tilde u(x,t)=\big[\bar{u}(x,t_{n-m+1}),\ldots,\bar{u}(x,t_n)\big]\in\mathbb{R}^{m d_u},\qquad t\in[0,T_{\mathrm{out}}].
    \]
    We augment $\tilde u$ with the space--time coordinates, concatenating $x\in\mathbb{R}^d$ and $t$, which yields an $\mathbb{R}^{m d_u+d+1}$-valued function on $\Omega\times[0,T_{\mathrm{out}}]$. A space--time FNO is then trained to predict the entire future slab on $\Omega\times[0,T_{\mathrm{out}}]$ in a single forward pass. For one-dimensional PDEs, this corresponds to the standard two-dimensional FNO implementation over space and time.
\end{itemize}

Both baselines use the same resolved history as input, but they expose temporal information to the network in a different way. \texttt{FNO1D} collapses the history into channels, while \texttt{FNO2D} repeats the same history tensor across the output window and relies on space--time Fourier mixing after this lifting. In contrast, our model applies a causal temporal kernel to the ordered history, allowing interactions among past time lags to be learned directly inside the history-to-future map. Apart from replacing the temporal-history representation, we keep the spatial Fourier mixing, conservative output parameterization, loss functions, and training configurations aligned as closely as possible, so that the comparison isolates the effect of temporal history mixing.

Table~\ref{tab:fno1d2d} shows that all three learned models reproduce the reference low-order statistics reasonably well: $\ell_{\mathrm{int}}$ and $\tau_E$ are close across the board, with \texttt{FNO2D} slightly overestimating $\ell_{\mathrm{int}}$ and \texttt{FNO1D} giving the closest $\tau_E$. The clearest separation appears in the energy-spectrum error $\varepsilon_S$, where the proposed spatiotemporal operator achieves the lowest value. This suggests that explicit temporal mixing over the resolved history improves multiscale statistical fidelity beyond what is captured by integral length and time scales alone. \texttt{FNO1D} has a parameter count comparable to our model but a larger $\varepsilon_S$, while \texttt{FNO2D} improves $\varepsilon_S$ at the cost of a much larger model. Computationally, \texttt{FNO2D} is fastest per epoch in our implementation; our model incurs a moderate overhead but remains near-\texttt{FNO1D} in parameter budget while delivering the best spectral accuracy. Overall, these results indicate that temporal history mixing improves spectral fidelity while retaining a compact parameter budget, supporting its role as a useful structural component for coarse-grained non-Markovian dynamics.

\begin{table}[t]
\footnotesize
\caption{Effect of temporal history representation on filtered KS dynamics: The integral length scale $\ell_{\mathrm{int}}$, the energy integral time scale $\tau_E$, the energy spectrum error $\varepsilon_S$, the number of trainable parameters, and the models' training cost.}
\label{tab:fno1d2d}
\begin{center}
\setlength{\tabcolsep}{4pt} 
\renewcommand{\arraystretch}{1.2}

  \begin{tabular}{|c|c|c|c|c|c|} \hline
   
   & $\ell_{\mathrm{int}}$
   & $\tau_E$
   & $\varepsilon_S (\downarrow)$
   & \bfseries \thead{Parameter \\counts $(\downarrow)$}
   & \bfseries \thead{Run time $(\downarrow)$\\ (s/epoch)} \\ \hline
    \texttt{FNO1D} & $1.730$ & $\mathbf{3.206}$ & $0.1422$ &$\mathbf{205,186}$ &$9.209$\\ 
    \texttt{FNO2D} & $1.757$ & $3.320$ & $0.09957$ &$3,596,738$ &$\mathbf{1.434}$ \\ 
    \texttt{\thead{Causal Spatiotemporal \\ neural operator}} & $\mathbf{1.728}$ & $3.238$ & $\mathbf{0.08465}$ &$206,341$ &$3.904$ \\  \hline
    \texttt{Reference} & $1.729$ & $3.112$ & \diagbox[dir=SW, width=5em, height=2em]{}{} & \diagbox[dir=SW, width=6em, height=2em]{}{} & \diagbox[dir=SW, width=8em, height=2em]{}{} \\ 
    
    \hline
  \end{tabular}
\end{center}
\end{table}

\section{Conclusions}
\label{sec:discussions}
We proposed a closure-free learning framework for coarse-grained multiscale PDE dynamics in which the resolved evolution is modeled as a windowed operator mapping a history slab on $\Omega\times[-T_{\mathrm{in}},0]$ to a future slab on $\Omega\times[0,T_{\mathrm{out}}]$. The architecture combines Fourier mixing in space with a causal temporal kernel that enforces a history-to-future directionality, and it incorporates a flux-form parameterization by predicting an effective flux and anchoring the conservative update at the last resolved snapshot. We further introduced a data-driven procedure to estimate the memory length based on the decorrelation time of an effective closure-injection diagnostic. On viscous Burgers, Kuramoto--Sivashinsky, and Navier--Stokes benchmarks, the resulting predictor produces stable autoregressive rollouts, reproduces key long-time statistics, and improves spectral and statistical fidelity over unconstrained, globally conservative, and baseline operator-learning models.

Several directions remain open. The present framework is deterministic and finite-memory, and can be interpreted as learning an effective resolved evolution conditioned on the recent history, rather than recovering a particular realization of the orthogonal-dynamics fluctuation term in the Mori--Zwanzig representation. A natural next step is to augment the windowed operator with stochastic components that explicitly represent unresolved fluctuation effects, especially in regimes without clear scale separation, while preserving the conservative flux-form structure. Adaptive or multiscale memory representations that extend beyond a fixed recent horizon, without sacrificing tractability in long autoregressive rollouts, are also worth developing. Whether the flux-form parameterization and causal temporal history mixing carry over to three-dimensional turbulence, complex geometries and boundary conditions, and more general coarse-graining operators remains to be tested.

\bibliographystyle{model1-num-names}
\bibliography{bib}
\end{document}